%% file: ldpcIEEE8.tex
\documentclass{article}
\usepackage{ITW06_PaperLaTeXStyle} 
\usepackage{amsmath}
\usepackage{amssymb}
\usepackage{epsfig}
\usepackage{color,multicol,graphics,pst-grad}

\newcommand{\C}{\mathcal{C}}
\renewcommand{\P}{\mathbb{P}}
\newcommand{\N}{\mathcal{N}}
\newcommand{\E}{\mathbb{E}}

\newcommand{\vers}{\rightarrow}

\definecolor{white}{rgb}{1,1,1}

\begin{document}
%\itwAfour             % Uncomment for A4 paper
\topmargin = 0mm

\itwtitle{Error Exponents of Low-Density Parity-Check Codes on the Binary Erasure Channel}

\itwauthor{Thierry Mora}
{Laboratoire de Physique Th\'eorique et Mod\`eles
Statistiques, B\^at.~100 \\ Universit\'e Paris-Sud and CNRS\\ F--91405 Orsay,
France.\\
Email: mora@lptms.u-psud.fr}

\itwsecondauthor{Olivier Rivoire}
{
Laboratory of Living Matter\\ The Rockefeller
University\\ 1230 York Avenue, Box 34\\ New York, NY--10021, USA \\
Email: orivoire@rockefeller.edu}

\itwmaketitle

\begin{itwabstract}
{\small We introduce a thermodynamic (large deviation) formalism for computing error exponents in error-correcting codes. Within this framework, we apply the heuristic cavity method from statistical mechanics to derive the average and typical error exponents of low-density parity-check (LDPC) codes on the binary erasure channel (BEC) under maximum-likelihood decoding.}
\end{itwabstract}

\begin{itwpaper}

\itwsection{Introduction}

Assessing the performance of error-correcting codes is a founding topics of information theory.
Amongst the simplest codes are the binary {\it block codes}, where a source generates with equal probability one of $2^L$ {\it codewords}, each a sequence of $N$ bits. As a codeword is transmitted through a {\it discrete memoryless channel}, a noise $\xi$ alters independently each bit with some probability. The {\it binary erasure channel} (BEC), for instance, erases a bit with a prescribed probability $p\in [0,1]$. Given the received message, the decoding task consists in inferring the most likely original codeword. The probability of error $\P_\xi({\rm error}|\C_N)$ then provides a simple characterization of the performance of a code $\C_N$.

The properties of error-correcting codes are conveniently studied through {\it ensembles} of codes $C_N$, consisting for instance of the set of all block codes with length $N$ and {\it rate} $R=L/N$. Shannon showed that, in the limit $N\to\infty$, a typical code in such an ensemble has a vanishing probability of error if (and only if) $R<R_c(p)$, where $R_c(p)$ corresponds to the {\it channel capacity}. This capacity is simply $R_c(p)=1-p$ for the BEC.
We are here interested in refining the description of the error probability beyond the channel capacity. {\it Error exponents} give the exponential rate of decay of $\P_\xi({\rm error}|\C_N)$ with $N$, for $\C_N\in C_N$, and offer the most appealing generalization. Of particular interest is the so-called {\it reliability function}, which gives the lowest achievable exponents as a function of the rate $R$ \cite{Berlekamp02}. However, despite significant efforts to estimate error exponents, resulting in the establishment of a number of bounds, exact expressions are scarce and restricted to a few extreme cases.

In this note, we put forward a {\it thermodynamic} (or {\it large deviation}) formalism~\cite{Ruelle04} for evaluating error exponents in error-correcting codes. This formalism coherently encompasses two types of exponents: if $C=\{C_N\}_{N\geq 1}$ denotes a sequence of ensembles of codes, we can indeed define, depending on the procedure for choosing the codes $\C_N$ in the ensembles $C_N$, an {\it average} and a {\it typical} error exponents as
\begin{eqnarray}
E_{\rm av}&=&-\lim_{N\vers \infty}\frac{1}{N}\log \E_{\C_N}\left[\P_\xi({\rm error}|\C_N)\right],\\
E_{\rm typ}&=&-\lim_{N\vers \infty}\frac{1}{N}\E_{\C_N}\left[\log \P_\xi({\rm error}|\C_N)\right],
\end{eqnarray}
where $\E_{\C_N}$ denotes the expectation value when $\C_N$ is drawn uniformly from the ensemble $C_N$ ($\log$ is base 2 throughout). Although the typical error exponent is the most interesting from the practical point of view, the average error exponent is usually simpler to estimate theoretically.

We analyze in the thermodynamic formalism one of the most promising family of block codes, the low-density parity-check (LDPC) codes~\cite{Gallager62}. The codewords of these codes correspond to the kernel of a sparse $M\times N$ {\it parity-check matrix} $A$, with $M=N-L$. Different choices for $A$ lead to different ensemble of codes $C_N$, the simplest example being regular ensembles\footnote{In this paper we restrict to regular codes, even though our method can be generalized to any irregular ensemble \cite{MoraRivoire06}.} defined with $A$ having $\ell$ 1's per column and $k$ per line, and zeros otherwise (in which case $R=1-\ell/k$). LDPC codes have been shown to formally map to physical models of disordered systems on random graphs~\cite{KabashimaSaad04}, and we shall exploit this analogy to apply the (non-rigorous) cavity method~\cite{Rivoire05} recently proposed in this context\footnote{While the exponential scaling of the error probability is guaranteed when the ensemble of codes comprises all block codes, the average error probability of LDPC codes is known to be polynomial in $N$~\cite{Gallager62}. Following Gallager, we shall ignore the few atypical codes responsible for this behavior, and consider the average error exponent associated with an expurgated ensemble where they have been excluded~\cite{Gallager62}.} (see also~\cite{SkantzosvanMourik03} for a related approach).

\itwsection{Thermodynamic formalism}

Given a received word, consisting of a codeword from a code $\C_N$ altered by a noise $\xi$ on the BEC, let $\N_N(\xi,\C_N)$ be the number of codewords from which it could come from (this quantity is independent of the initial codeword with LDPC codes).  By definition, decoding is achievable if and only if $\N_N(\xi,\C_N)=1$. For random codes, the geometry of the space of codewords indicates that, at least in the vicinity of the channel capacity, an error most probably involves an exponential number of potential codewords (see e.g.~\cite{BargForney02}). In such situations, we characterize $\N_N(\xi,\C_N)$ by an {\it entropy}, defined as
\begin{equation}
S_N(\xi,\C_N)=\log \N_N(\xi,\C_N).
\end{equation}
In the limit $N\to\infty$, for sequences of codes $\C=\{\C_N\}_N $ taken from the sequence of ensembles $C=\{C_N\}_N$, the entropy density $s=S_N/N$ concentrates to a well defined value $\bar s$, and the channel coding theorem takes the following form: there exists $p_c$, such that $\bar s=0$ for $p<p_c$, and $\bar s>0$ for $p>p_c$~\cite{FranzLeone02}. More generally, we postulate that, for a typical sequence of codes $\C^0=\{\C^0_N\}_N$, the entropy $S_N$ satisfies a {\it large deviation principle}~\cite{denHollander00}, i.e.,
\begin{equation}\label{eq:Lc}
\P_\xi[S_N(\xi,\C^0_N)/N=s]\asymp 2^{-NL_0(s)},
\end{equation}
with $a_N\asymp b_N$ meaning that $\log a_N/\log b_N\vers 1$. The typical value $\bar s$ corresponds here to the minimum of the {\it rate function} $L_0$, with $L_0(\bar s)=0$.
In cases where $L_0$ is strictly convex, the typical error exponent is obtained as
\begin{equation}
\begin{split}
E_{\rm typ} &=-\lim_{N\vers\infty}\frac{1}{N}\log \sum_{s\geq 1/N}\P_\xi[S_N(\xi,\C^0_N)/N=s]\\
&=L_0(s=0).
\end{split}
\end{equation}
A simpler quantity to compute than $L_0(s)$ is $L_1(s)$, the rate function for the large deviations of $S_N(\xi,\C_N)$ with respect to both the noise $\xi$ and the codes $\C_N$,
\begin{equation}
\P_{\xi,\C_N}[S_N(\xi,\C_N)/N=s]\asymp 2^{-NL_1(s)}.
\end{equation}
In the so-called {\it thermodynamic formalism}~\cite{Ruelle04}, $L_1(s)$ is associated with a {\it potential} $\phi(x)$ defined through the relation
\begin{equation}\label{eq:phiC}
2^{N\phi(x)}=\E_{\xi,\C_N}[2^{xS_N(\xi,\C_N)}]\asymp\int {\rm d}s\ 2^{N[xs-L_1(s)]}.
\end{equation}
Under the assumption that it is convex, the rate function $L_1(s)$ is derived from the knowledge of $\phi(x)$ by Legendre transformation:
\begin{equation}
L_1(s)=\max_x\left[xs-\phi(x)\right].
\end{equation}
The average exponent, obtained from $E_{\rm av}=L_1(s=0)$, may differ from the typical exponent $E_{\rm typ}$. Typical codes $\C_N^0$ can however also be described within a thermodynamic formalism, provided an extra ``temperature'' $y$ is introduced, together with a generalized potential $\psi(x,y)$ satisfying
\begin{equation}\label{eq:defpsi}
2^{N\psi(x,y)}=\E_{\C_N}\left[\left(\E_\xi[2^{xS_N(\xi,\C_N)}]\right)^y\right].
\end{equation}
The average case is here recovered for $y=1$, with $\psi(x,y=1)=\phi(x)$. Typical error exponents are associated with $y=0$ (see~\cite{MoraRivoire06} for details and exceptions), with
\begin{equation}
E_{\rm typ}=L_0(s=0)=-\partial_y \psi(x^*,y=0),
\end{equation}
where $x^*$ selects for $s=\left.\frac{1}{y}\partial_x \psi(x^*,y)\right|_{y=0}=0$.

\itwsection{Cavity method}

Disordered systems constructed out of random ensembles, of which LDPC codes are particular examples, have been the subject of intensive studies in statistical mechanics. One of the most elaborate analytical tool developed in this context is the {\it cavity method}~\cite{MezardParisi01}, which allows to extract the typical properties of models defined on random graphs. While yielding virtually equivalent predictions than the similar {\it replica method}, this method has both more sound probabilistic foundations, and an attractive relation to message-passing algorithms, such as {\it belief propagation} (BP). The cavity method has also been recently extended to deal with large deviations~\cite{Rivoire05}, making it perfectly suited to the evaluation of error exponents.

As far as typical codes and typical noise are concerned, the cavity method is equivalent to a BP {\it density evolution} analysis. Belief propagation, also known as the ``peeling decoder'' in the context of the BEC \cite{LubyMitzenmacher01}, consists in propagating messages between {\it bits} (the $N$ letters of a word) and {\it checks} (the $M$ linear equations encoded in the parity-check matrix $A$ that each codeword must satisfy). The messages can take three different values: $*$ (erasure) or $0$ or $1$. Initially, each bit sends its value 0 or 1, or $*$ if erased, to each of the parity checks it is involved in. Check-to-bit and bit-to-check messages are then sent alternatively. If a check $a$ receives non-erasure messages from all its bits but $i$, it sends to $i$ the sum (modulo $2$) of these messages; otherwise, the check $a$ sends $*$ to $i$. If an erased bit $i$ receives at least one non-erasure message from any of its checks but $a$, it sends it to $a$ (if more than one, they are necessarily identical); otherwise, the bit $i$ sends its value, 0 or 1, or $*$ if erased, to $a$. The algorithm stops after convergence of the iterations.

The (typical) cavity method, or BP density evolution, analyzes the outcome of this procedure in the limit where the codeword length $N$ is infinite. It introduces $\eta$, the probability that a bit sends an erasure message to a check, and $\zeta$ the probability that a check sends an erasure message to a bit, both taken after BP has reached convergence. The {\it cavity equations} satisfied by these two probabilities,
\begin{equation}\label{eq:cav}
\zeta =1-(1-\eta)^{k-1},\qquad
\eta = p\zeta^{\ell-1},
\end{equation}
characterize the fixed point of the BP density evolution (see Fig.~\ref{fig:cavity}).

\begin{figure}[h]
\begin{center}
\resizebox{.85\linewidth}{!}{\input{cavity.pstex_t}}
\caption{Illustration of the cavity equations \eqref{eq:cav}, with $k=4$ and $\ell=3$. (a): a check node (square) sends an erasure message to a bit node (dashed circle) if at least one of its other variables sends an erasure message. (b): a bit node (circle) sends an erasure message to a check node (dashed square) if it has been erased and if all its other checks send an erasure message.}
\label{fig:cavity}
\end{center}
\end{figure}
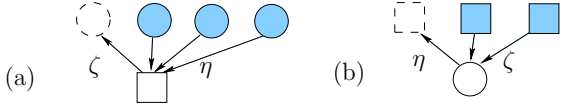

Once BP has converged, bits receiving at least one non-erasure message are fixed to their correct value, as are the non-erased bits. When eliminated, along with the checks receiving no more than one erasure message, they leave the so-called {\it core}. The dimensions $M_c\times N_c$ of the associated residual matrix are, with high probability:
\begin{equation}
\begin{split}
N_c&= p\zeta^\ell N+o(N),\\
M_c&= \frac{\ell}{k}[1-(1-\eta)^k-k\eta(1-\eta)^{k-1}]N+o(N).
\end{split}
\end{equation}
For $p<p_d(\ell,k)$, the only solution to \eqref{eq:cav} is $\zeta=0,\eta=0$, meaning that BP is able to decode the whole word with high probability. For $p>p_d$ however, BP gets stuck at some $\zeta>0$, $\eta>0$. In this case, it can be proved that the residual matrix has full-rank with high probability \cite{MeassonMontanari04}. Therefore, the problem has exactly $2^{N_c-M_c}$ solutions if $N_c>M_c$, and one solution (the original codeword) otherwise. In this approach, the critical noise $p_c(\ell,k)$ is obtained from the condition $N_c=M_c$, and $\bar s$ is given by $\max(0,\bar s_{\rm cav})$, with $\bar s_{\rm cav}=\lim_{N\vers\infty} (N_c-M_c)/N$.

The large deviation cavity method is built on the same ideas but incorporates a biased measure over the noise and code ensemble, as prescribed by Eq.~\eqref{eq:defpsi}. When we consider the value of a bit-to-check message as a function of its $(\ell-1)(k-1)$ ``grandparents'', we also evaluate the ``entropy shift'' $\Delta S$ associated with the addition of the bit and its $\ell-1$ checks, i.e. the difference between the numbers of columns and lines contributed by the bit and its checks to the residual matrix. Then the message is sent with a probability proportional to
\begin{equation}
\left(\E_\xi 2^{x\Delta S}\right)^y.
\end{equation}
For regular LDPC codes, we thus obtain for the potential
\begin{equation}
\begin{split}
&\psi(x,y)=\\
&\log Z_\ell-\frac{\ell(k-1)}{k}\log\left[(1-\eta)^k+(1-(1-\eta)^k)2^{-xy}\right]
\end{split}
\end{equation}
with
\begin{equation}
Z_\ell=(\zeta 2^{-xy}+1-\zeta)^\ell-(\zeta 2^{-xy})^\ell+\zeta^\ell(p2^x+1-p)^y2^{-\ell xy}
\end{equation}
and
\begin{equation}\label{eq:cavldcm}
\begin{split}
\eta &=\zeta^{\ell-1}(p2^x)^y2^{-(\ell-1)xy} Z_{\ell-1}^{-1},\\
\zeta &=1-(1-\eta)^{k-1}.
\end{split}
\end{equation}
Note that the entropy conjugated with $x$ is not the ``real'' entropy $s$, but $s_{\rm cav}=(N_c-M_c)/N$.
When $x=0$, the fixed point of the usual density evolution equations~\eqref{eq:cav} is recovered, with $(1/y)\partial_x \psi({x=0,y})$ giving back $\bar s_{\rm cav}$, the typical value.\\

\ \\

\itwsection{LDPC codes}

We first discuss average error exponents.
The calculation of the average rate function $L_1(s)$ reveals four distinct regimes when the noise level $p$ is varied, as illustrated and explained in Fig.~\ref{largedevbec}. In particular, we find that the rate function $L_1(s)$ is no longer defined for $s=0$ when $p$ is too small ($p<p_{\rm 1rsb}$), which points to the inadequacy of our method in this low-noise regime. 

\begin{figure}
\begin{center}
\resizebox{.49\linewidth}{!}{\input{largedevbec1}}\hfill
\resizebox{.49\linewidth}{!}{\input{largedevbec2}}
\resizebox{.49\linewidth}{!}{\input{largedevbec3}}\hfill
\resizebox{.49\linewidth}{!}{\input{largedevbec4}}
\caption{Average entropic rate function $L_1(s)$ as a function of the entropy density $s_{\rm cav}$, for the regular LDPC code $\ell=3$, $k=6$ on the BEC with increasing values of $p$. 
The real entropy is actually $s=\max(0,s_{\rm cav})$.
(a): $p<p_{\rm 1rsb}$, no solution with $s=0$; (b): $p_{\rm 1rsb}
<p<p_d$, a solution with $s=0$, but $\bar s$ is not defined; (c): $p_d<p<p_c$, $\bar s=0$; (d): $p>p_c$, $\bar s>0$ indicates that decoding typically fails.}
\label{largedevbec}
\end{center}
\end{figure}
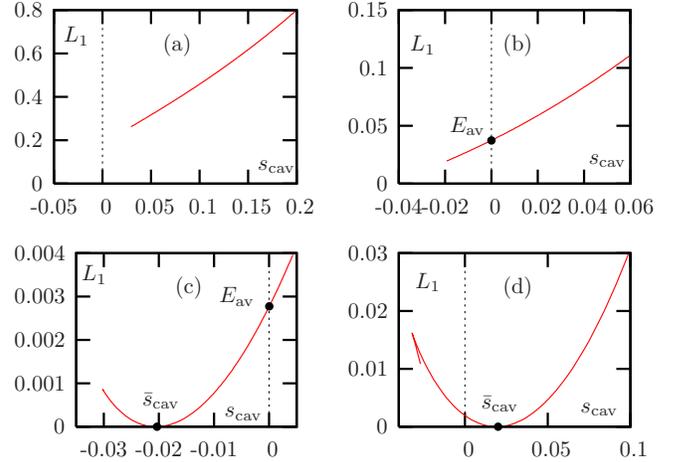

Indeed, by retaining $s=0$ as criterion for correct decoding, we assumed that an error implicates an exponential number of codewords. An error may however also be caused by the presence of one (or a few) isolated codeword(s). Estimating this probability requires an alternative, ``energetic'', scheme, as opposed to the ``entropic'' scheme discussed so far\footnote{The energetic version of the cavity method is also referred to as ``replica symmetric'' in the physics literature, while the entropic version is known as ``one-step replica symmetry breaking''.}. Equations for the energetic average and typical error exponents can also be obtained from the large deviation cavity method~\cite{MoraRivoire06}, but their solutions are confined to a restricted interval $p>p_{\rm rs}$, indicating again that the lowest noise levels are not appropriately described. The entropic and energetic exponents are found to cross at $p_e$, which corresponds to the so-called {\em critical rate}~\cite{BargForney02,Gallager68}. We conjecture that the entropic exponent, as given by the above equations, is exact in the range $[p_e,p_c]$, while the energetic exponent (not presented here), which applies for $[p_{\rm rs},p_e]$, is only approximate.

Fig.~\ref{becerrexp} shows our predictions for the average exponent of the $\ell=3$, $k=6$ regular LDPC codes, with the two regimes represented; the same general picture holds for other regular or irregular ensembles (see also Table~\ref{noiselevels}).

\begin{figure}
\begin{center}
\resizebox{\linewidth}{!}{\input{errexpldpc}}
\caption{Average error exponent as
a function of the noise level $p$ of the BEC for the regular LDPC code ensemble
with $k=6$ and $\ell=3$. Gallager's union bound  and
the random linear code limit \eqref{eq:RLMav} are also plotted for comparison.}
\label{becerrexp}
\end{center}
\end{figure}
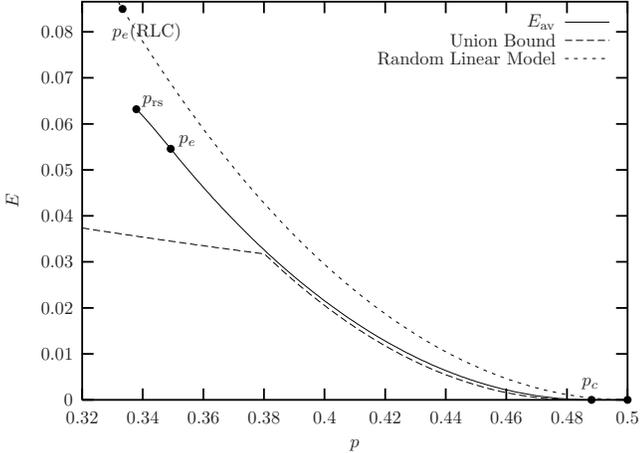

\begin{table}
\begin{center}
\begin{tabular}{|l|l|l|}
\hline 
$(k,\ell)$ &		$(4,3)$  	&$(6,3)$ \\ \hline
$p_{\textrm{1rsb}}$ &	$0.3252629709$  &$0.2668568754$\\ \hline
$p_{\rm rs}$ &		$0.5465748811$ 	&$0.3378374641$\\ \hline
$p_e$ &			$0.6068720166$ 	&$0.3491884902$\\ \hline
$p_d$ &			$0.6474256494$	&$0.4294398144$\\ \hline
$p_c$ &			 $0.7460097025$	&$0.4881508842$\\ \hline
\end{tabular}
\caption{ Thresholds
$p_{\textrm{1rsb}}$, $p_{\rm rs}$, $p_e$, $p_d$ and $p_c$ (see text and Fig.~\ref{largedevbec}) for two regular ensembles of LDPC codes.}
\label{noiselevels}
\end{center}
\end{table}

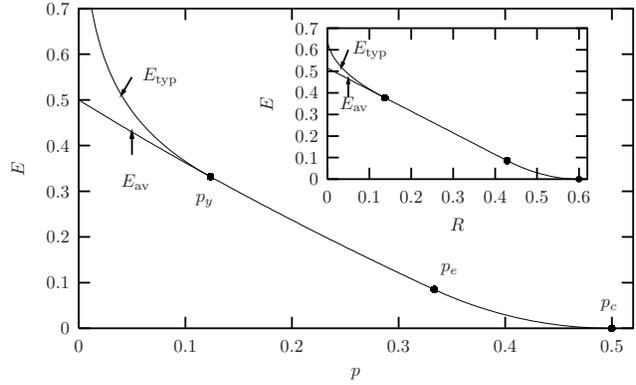
\begin{figure}
\begin{center}
\resizebox{\linewidth}{!}{\input{rlmbec}}
\caption{Average and typical error exponents of random linear codes on the BEC as a function of $p$, with $R=1/2$ fixed. Inset: the same exponents as a function of $R$, with $p=0.4$ fixed.}
\label{rlmbec}
\end{center}
\end{figure}

\itwsection{The Random Linear Code limit}

This limit is obtained from regular codes with $k,\ell\vers\infty$ and $R=1-\ell/k$ fixed, where the potential simplifies to:
\begin{equation}
\psi(x,y)=y\log(p2^x+1-p)+(R-1) xy.
\end{equation}
The trivial dependence of $\psi(x,y)$ with $y$ implies that the two error exponents $E_{\rm av}$ and $E_{\rm typ}$, as obtained from the entropic scheme, are identical. They are equal to the {\it volume bound} \cite{Berlekamp02} $D(1-R||p)$, where $D(x||y)=x\log(x/y)+(1-x)\log((1-x)/(1-y))$ denotes the {\it Kullback-Leibler divergence}.

The intersection of the entropic and energetic average error exponents yields the threshold
\begin{equation}
p_e=\frac{1-R}{1+R},
\end{equation}
and we obtain for the average error exponent in the infinite connectivity limit:
\begin{equation}\label{eq:RLMav}
E_{\rm av}(\textrm{RLC})=\left\{\begin{array}{ll}1-R-\log(1+p) 
& \textrm{if }p<p_e,\\
D(1-R||p)&\textrm{if }p_e<p<p_c.\end{array}\right.
\end{equation}
It coincides with the average error exponent of the {\it random linear code} (RLC) ensemble, where the $M\times N$ parity-check matrix is chosen at random with uniform probability among all possible parity-check matrices. Assuming that the inversion of the limits $N\vers \infty$ and $k,\ell\vers\infty$ is justified, we interpret this result as a validation of our approach (note that here, $p_{\rm rs}=0$).

The analysis of the typical error exponent in the energetic regime leads us to introduce an additional threshold,
\begin{equation}
p_y=\frac{\delta_{GV}(R)}{1-\delta_{GV}(R)},
\end{equation}
where $\delta_{GV}(R)$, the minimal reduced distance of a typical linear code~\cite{BargForney02}, is given by the smallest solution of $-\delta\log \delta-(1-\delta)\log(1-\delta)=1-R$. Below $p_y$, physical arguments~\cite{MoraRivoire06} indicates that the typical error exponent must differ from the average one, with:

\begin{equation}\label{eq:RLMtyp}
E_{\rm typ}(\textrm{RLC})=\left\{
\begin{array}{ll}-\delta_{GV}(R)\log p & \textrm{if }p<p_y,\\
E_{\rm av}(\textrm{RLC})&\textrm{if }p>p_y.\end{array}\right.
\end{equation}
 We are not aware of any previous report of this expression in the literature, but the fact that it matches the {\it union bound} suggests that it is exact. Fig.~\ref{rlmbec} presents the error exponents as a function of $p$ for a fixed value of the rate $R=1/2$.

The two thresholds $p_e$ and $p_y$ are presumably generic features of block codes, and are also found with random codes on the binary symmetric channel~\cite{BargForney02}.\\
\ \\

\itwsection{Discussion}

Despite being one of the earliest and most basic topics in information theory, error exponents still retain today a number of unsolved issues. We advocated here a novel, thermodynamical, formulation of this problem. Using the cavity method from statistical mechanics, we worked out in this framework expressions for the average and typical error exponents of LDPC codes on the BEC. 

While non rigorous, the cavity method aims at providing exact formul\ae. Accordingly, our expressions are consistent with the various rigorous studies reported in the literature. The quest for rigorous proofs of formul\ae\ obtained from the cavity method is currently an active fields of mathematics~\cite{Talagrand03}. Remarkably, predictions from the cavity method on the maximum-likelihood threshold $p_c$ \cite{FranzLeone02} could be turned into rigorous theorems \cite{MeassonMontanari04}.
This may inspire alternative derivations of our results. 

Perhaps not too surprisingly, the entropic range $p_e<p<p_c$ where we conjecture our results to be exact also coincides with the limited interval for which the related problem of determining the reliability function of block codes has been solved so far. Extending our method to $p<p_e$, where we could obtain only approximate results (except in the infinite connectivity limit), remains a challenging open problem.

Using the same approach, we also analyzed the case of the binary symmetric channel, obtaining comparable results~\cite{MoraRivoire06}. A more interesting extension would be to iterative decoding, such as BP. Although arguably quite academic, studying maximum-likelihood decoding, as we did, is nevertheless certainly an essential preliminary step.

\itwacknowledgments

It is a pleasure to thank Stefano Ciliberti, Marc M\'ezard and Lenka Zdeborov{\'a} for their critical reading.
The work of T.M. was supported in part by the EC through the network
MTR 2002-00319 `STIPCO' and the FP6 IST consortium `EVERGROW'. O.R. is a fellow of the Human Frontier Science Program.

\end{itwpaper}

\begin{itwreferences}

\bibitem{BargForney02}
A.~Barg and G.~D.~Forney Jr.,
 ``Random codes : minimum distances and error exponents,''
 {\em IEEE Trans. Inform. Theory}, 48:2568--2573, 2002.

\bibitem{Berlekamp02}
E.~R. Berlekamp,
 ``The performance of block codes,''
 {\em Notices of the AMS}, pages 17--22, January 2002.

\bibitem{denHollander00}
F.~den Hollander,
 {\em Large deviations},
 Fields Institute Monographs 14. American Mathematical Society,
  Providence RI, 2000.

\bibitem{FranzLeone02}
S.~Franz, M.~Leone, A.~Montanari, and F.~Ricci-Tersenghi,
 ``The dynamic phase transition for decoding algorithms,''
 {\em Phys. Rev. E}, 66:046120, 2002.

\bibitem{Gallager62}
R.~G. Gallager,
 ``Low-density parity check codes,''
 {\em IRE Trans. Inf. Theory}, IT-8:21, 1962.

\bibitem{Gallager68}
R.~G. Gallager,
 {\em Information theory and reliable communication},
 John Wiley and Sons, New York, 1968.

\bibitem{KabashimaSaad04}
Y.~Kabashima and D.~Saad,
 ``Statistical mechanics of low-density parity-check codes,''
 {\em J. Phys. A: Math. Gen}, 37:R1--R43, 2004.

\bibitem{LubyMitzenmacher01}
M.~Luby, M.~Mitzenmacher, A.~Shokrollahi, and D.~Spielman, ``Efficient erasure correcting codes,'' {\em IEEE Trans. Inform. Theory}, vol. 47, 569--584, Feb. 2001.

\bibitem{MeassonMontanari04}
C.~Measson, A.~Montanari, T.~Richardson, and R.~Urbanke,
 ``Life above threshold: from list decoding to area theorem and {MSE},''
 In {\em Proc. ITW}, San Antonio, USA, October 2004.

\bibitem{MezardParisi01}
M.~M{\'e}zard and G.~Parisi.
 ``The Bethe lattice spin glass revisited,''
 {\em Eur. Phys. J. B}, 20:217, 2001.

\bibitem{MoraRivoire06}
T.~Mora and O.~Rivoire,
 2006.
 In preparation.

\bibitem{Rivoire05}
O.~Rivoire.
 ``The cavity method for large deviations,''
 {\em J. Stat. Mech.}, P07004, 2005.

\bibitem{Ruelle04}
D. Ruelle.
{\em Thermodynamic formalism}, Cambridge Math. Library,
2nd Ed, 2004.

\bibitem{SkantzosvanMourik03}
N.~S. Skantzos, J.~van Mourik, D.~Saad, and Y.~Kabashima,
 ``Average and reliability error exponents in low-density parity-check
  codes,''
 {\em J. Phys. A}, 36:11131--11141, 2003.

\bibitem{Talagrand03}
M.~Talagrand,
{\em Spin glasses : a challenge for mathematicians. Cavity and mean field models},
Springer-Verlag,
New-York,
2003.

\end{itwreferences}

\end{document}

%% file: cavity.pstex_t
\begin{picture}(0,0)%
\includegraphics{cavity.pstex}%
\end{picture}%
\setlength{\unitlength}{4144sp}%
\begingroup\makeatletter\ifx\SetFigFont\undefined%
\gdef\SetFigFont#1#2#3#4#5{%
  \reset@font\fontsize{#1}{#2pt}%
  \fontfamily{#3}\fontseries{#4}\fontshape{#5}%
  \selectfont}%
\fi\endgroup%
\begin{picture}(4209,780)(5833,-1693)
\put(7309,-1459){\makebox(0,0)[lb]{\smash{{\SetFigFont{12}{14.4}{\rmdefault}{\mddefault}{\updefault}{\color[rgb]{0,0,0}$\eta$}%
}}}}
\put(5833,-1549){\makebox(0,0)[lb]{\smash{{\SetFigFont{12}{14.4}{\rmdefault}{\mddefault}{\updefault}{\color[rgb]{0,0,0}(a)}%
}}}}
\put(8317,-1513){\makebox(0,0)[lb]{\smash{{\SetFigFont{12}{14.4}{\rmdefault}{\mddefault}{\updefault}{\color[rgb]{0,0,0}(b)}%
}}}}
\put(9605,-1420){\makebox(0,0)[lb]{\smash{{\SetFigFont{12}{14.4}{\rmdefault}{\mddefault}{\updefault}{\color[rgb]{0,0,0}$\zeta$}%
}}}}
\put(6469,-1459){\makebox(0,0)[lb]{\smash{{\SetFigFont{12}{14.4}{\rmdefault}{\mddefault}{\updefault}{\color[rgb]{0,0,0}$\zeta$}%
}}}}
\put(8921,-1408){\makebox(0,0)[lb]{\smash{{\SetFigFont{12}{14.4}{\rmdefault}{\mddefault}{\updefault}{\color[rgb]{0,0,0}$\eta$}%
}}}}
\end{picture}%

%% file: largedevbec1.tex
% GNUPLOT: LaTeX picture with Postscript
\begingroup%
  \makeatletter%
  \newcommand{\GNUPLOTspecial}{%
    \@sanitize\catcode`\%=14\relax\special}%
  \setlength{\unitlength}{0.1bp}%
\begin{picture}(1440,1080)(0,0)%
{\GNUPLOTspecial{"
%!PS-Adobe-2.0 EPSF-2.0
%%Title: largedevbec1.tex
%%Creator: gnuplot 4.0 patchlevel 0
%%CreationDate: Wed May 17 12:13:55 2006
%%DocumentFonts: 
%%BoundingBox: 0 0 144 108
%%Orientation: Landscape
%%EndComments
/gnudict 256 dict def
gnudict begin
/Color true def
/Solid false def
/gnulinewidth 5.000 def
/userlinewidth gnulinewidth def
/vshift -33 def
/dl {10.0 mul} def
/hpt_ 31.5 def
/vpt_ 31.5 def
/hpt hpt_ def
/vpt vpt_ def
/Rounded false def
/M {moveto} bind def
/L {lineto} bind def
/R {rmoveto} bind def
/V {rlineto} bind def
/N {newpath moveto} bind def
/C {setrgbcolor} bind def
/f {rlineto fill} bind def
/vpt2 vpt 2 mul def
/hpt2 hpt 2 mul def
/Lshow { currentpoint stroke M
  0 vshift R show } def
/Rshow { currentpoint stroke M
  dup stringwidth pop neg vshift R show } def
/Cshow { currentpoint stroke M
  dup stringwidth pop -2 div vshift R show } def
/UP { dup vpt_ mul /vpt exch def hpt_ mul /hpt exch def
  /hpt2 hpt 2 mul def /vpt2 vpt 2 mul def } def
/DL { Color {setrgbcolor Solid {pop []} if 0 setdash }
 {pop pop pop 0 setgray Solid {pop []} if 0 setdash} ifelse } def
/BL { stroke userlinewidth 2 mul setlinewidth
      Rounded { 1 setlinejoin 1 setlinecap } if } def
/AL { stroke userlinewidth 2 div setlinewidth
      Rounded { 1 setlinejoin 1 setlinecap } if } def
/UL { dup gnulinewidth mul /userlinewidth exch def
      dup 1 lt {pop 1} if 10 mul /udl exch def } def
/PL { stroke userlinewidth setlinewidth
      Rounded { 1 setlinejoin 1 setlinecap } if } def
/LTw { PL [] 1 setgray } def
/LTb { BL [] 0 0 0 DL } def
/LTa { AL [1 udl mul 2 udl mul] 0 setdash 0 0 0 setrgbcolor } def
/LT0 { PL [] 1 0 0 DL } def
/LT1 { PL [4 dl 2 dl] 0 1 0 DL } def
/LT2 { PL [2 dl 3 dl] 0 0 1 DL } def
/LT3 { PL [1 dl 1.5 dl] 1 0 1 DL } def
/LT4 { PL [5 dl 2 dl 1 dl 2 dl] 0 1 1 DL } def
/LT5 { PL [4 dl 3 dl 1 dl 3 dl] 1 1 0 DL } def
/LT6 { PL [2 dl 2 dl 2 dl 4 dl] 0 0 0 DL } def
/LT7 { PL [2 dl 2 dl 2 dl 2 dl 2 dl 4 dl] 1 0.3 0 DL } def
/LT8 { PL [2 dl 2 dl 2 dl 2 dl 2 dl 2 dl 2 dl 4 dl] 0.5 0.5 0.5 DL } def
/Pnt { stroke [] 0 setdash
   gsave 1 setlinecap M 0 0 V stroke grestore } def
/Dia { stroke [] 0 setdash 2 copy vpt add M
  hpt neg vpt neg V hpt vpt neg V
  hpt vpt V hpt neg vpt V closepath stroke
  Pnt } def
/Pls { stroke [] 0 setdash vpt sub M 0 vpt2 V
  currentpoint stroke M
  hpt neg vpt neg R hpt2 0 V stroke
  } def
/Box { stroke [] 0 setdash 2 copy exch hpt sub exch vpt add M
  0 vpt2 neg V hpt2 0 V 0 vpt2 V
  hpt2 neg 0 V closepath stroke
  Pnt } def
/Crs { stroke [] 0 setdash exch hpt sub exch vpt add M
  hpt2 vpt2 neg V currentpoint stroke M
  hpt2 neg 0 R hpt2 vpt2 V stroke } def
/TriU { stroke [] 0 setdash 2 copy vpt 1.12 mul add M
  hpt neg vpt -1.62 mul V
  hpt 2 mul 0 V
  hpt neg vpt 1.62 mul V closepath stroke
  Pnt  } def
/Star { 2 copy Pls Crs } def
/BoxF { stroke [] 0 setdash exch hpt sub exch vpt add M
  0 vpt2 neg V  hpt2 0 V  0 vpt2 V
  hpt2 neg 0 V  closepath fill } def
/TriUF { stroke [] 0 setdash vpt 1.12 mul add M
  hpt neg vpt -1.62 mul V
  hpt 2 mul 0 V
  hpt neg vpt 1.62 mul V closepath fill } def
/TriD { stroke [] 0 setdash 2 copy vpt 1.12 mul sub M
  hpt neg vpt 1.62 mul V
  hpt 2 mul 0 V
  hpt neg vpt -1.62 mul V closepath stroke
  Pnt  } def
/TriDF { stroke [] 0 setdash vpt 1.12 mul sub M
  hpt neg vpt 1.62 mul V
  hpt 2 mul 0 V
  hpt neg vpt -1.62 mul V closepath fill} def
/DiaF { stroke [] 0 setdash vpt add M
  hpt neg vpt neg V hpt vpt neg V
  hpt vpt V hpt neg vpt V closepath fill } def
/Pent { stroke [] 0 setdash 2 copy gsave
  translate 0 hpt M 4 {72 rotate 0 hpt L} repeat
  closepath stroke grestore Pnt } def
/PentF { stroke [] 0 setdash gsave
  translate 0 hpt M 4 {72 rotate 0 hpt L} repeat
  closepath fill grestore } def
/Circle { stroke [] 0 setdash 2 copy
  hpt 0 360 arc stroke Pnt } def
/CircleF { stroke [] 0 setdash hpt 0 360 arc fill } def
/C0 { BL [] 0 setdash 2 copy moveto vpt 90 450  arc } bind def
/C1 { BL [] 0 setdash 2 copy        moveto
       2 copy  vpt 0 90 arc closepath fill
               vpt 0 360 arc closepath } bind def
/C2 { BL [] 0 setdash 2 copy moveto
       2 copy  vpt 90 180 arc closepath fill
               vpt 0 360 arc closepath } bind def
/C3 { BL [] 0 setdash 2 copy moveto
       2 copy  vpt 0 180 arc closepath fill
               vpt 0 360 arc closepath } bind def
/C4 { BL [] 0 setdash 2 copy moveto
       2 copy  vpt 180 270 arc closepath fill
               vpt 0 360 arc closepath } bind def
/C5 { BL [] 0 setdash 2 copy moveto
       2 copy  vpt 0 90 arc
       2 copy moveto
       2 copy  vpt 180 270 arc closepath fill
               vpt 0 360 arc } bind def
/C6 { BL [] 0 setdash 2 copy moveto
      2 copy  vpt 90 270 arc closepath fill
              vpt 0 360 arc closepath } bind def
/C7 { BL [] 0 setdash 2 copy moveto
      2 copy  vpt 0 270 arc closepath fill
              vpt 0 360 arc closepath } bind def
/C8 { BL [] 0 setdash 2 copy moveto
      2 copy vpt 270 360 arc closepath fill
              vpt 0 360 arc closepath } bind def
/C9 { BL [] 0 setdash 2 copy moveto
      2 copy  vpt 270 450 arc closepath fill
              vpt 0 360 arc closepath } bind def
/C10 { BL [] 0 setdash 2 copy 2 copy moveto vpt 270 360 arc closepath fill
       2 copy moveto
       2 copy vpt 90 180 arc closepath fill
               vpt 0 360 arc closepath } bind def
/C11 { BL [] 0 setdash 2 copy moveto
       2 copy  vpt 0 180 arc closepath fill
       2 copy moveto
       2 copy  vpt 270 360 arc closepath fill
               vpt 0 360 arc closepath } bind def
/C12 { BL [] 0 setdash 2 copy moveto
       2 copy  vpt 180 360 arc closepath fill
               vpt 0 360 arc closepath } bind def
/C13 { BL [] 0 setdash  2 copy moveto
       2 copy  vpt 0 90 arc closepath fill
       2 copy moveto
       2 copy  vpt 180 360 arc closepath fill
               vpt 0 360 arc closepath } bind def
/C14 { BL [] 0 setdash 2 copy moveto
       2 copy  vpt 90 360 arc closepath fill
               vpt 0 360 arc } bind def
/C15 { BL [] 0 setdash 2 copy vpt 0 360 arc closepath fill
               vpt 0 360 arc closepath } bind def
/Rec   { newpath 4 2 roll moveto 1 index 0 rlineto 0 exch rlineto
       neg 0 rlineto closepath } bind def
/Square { dup Rec } bind def
/Bsquare { vpt sub exch vpt sub exch vpt2 Square } bind def
/S0 { BL [] 0 setdash 2 copy moveto 0 vpt rlineto BL Bsquare } bind def
/S1 { BL [] 0 setdash 2 copy vpt Square fill Bsquare } bind def
/S2 { BL [] 0 setdash 2 copy exch vpt sub exch vpt Square fill Bsquare } bind def
/S3 { BL [] 0 setdash 2 copy exch vpt sub exch vpt2 vpt Rec fill Bsquare } bind def
/S4 { BL [] 0 setdash 2 copy exch vpt sub exch vpt sub vpt Square fill Bsquare } bind def
/S5 { BL [] 0 setdash 2 copy 2 copy vpt Square fill
       exch vpt sub exch vpt sub vpt Square fill Bsquare } bind def
/S6 { BL [] 0 setdash 2 copy exch vpt sub exch vpt sub vpt vpt2 Rec fill Bsquare } bind def
/S7 { BL [] 0 setdash 2 copy exch vpt sub exch vpt sub vpt vpt2 Rec fill
       2 copy vpt Square fill
       Bsquare } bind def
/S8 { BL [] 0 setdash 2 copy vpt sub vpt Square fill Bsquare } bind def
/S9 { BL [] 0 setdash 2 copy vpt sub vpt vpt2 Rec fill Bsquare } bind def
/S10 { BL [] 0 setdash 2 copy vpt sub vpt Square fill 2 copy exch vpt sub exch vpt Square fill
       Bsquare } bind def
/S11 { BL [] 0 setdash 2 copy vpt sub vpt Square fill 2 copy exch vpt sub exch vpt2 vpt Rec fill
       Bsquare } bind def
/S12 { BL [] 0 setdash 2 copy exch vpt sub exch vpt sub vpt2 vpt Rec fill Bsquare } bind def
/S13 { BL [] 0 setdash 2 copy exch vpt sub exch vpt sub vpt2 vpt Rec fill
       2 copy vpt Square fill Bsquare } bind def
/S14 { BL [] 0 setdash 2 copy exch vpt sub exch vpt sub vpt2 vpt Rec fill
       2 copy exch vpt sub exch vpt Square fill Bsquare } bind def
/S15 { BL [] 0 setdash 2 copy Bsquare fill Bsquare } bind def
/D0 { gsave translate 45 rotate 0 0 S0 stroke grestore } bind def
/D1 { gsave translate 45 rotate 0 0 S1 stroke grestore } bind def
/D2 { gsave translate 45 rotate 0 0 S2 stroke grestore } bind def
/D3 { gsave translate 45 rotate 0 0 S3 stroke grestore } bind def
/D4 { gsave translate 45 rotate 0 0 S4 stroke grestore } bind def
/D5 { gsave translate 45 rotate 0 0 S5 stroke grestore } bind def
/D6 { gsave translate 45 rotate 0 0 S6 stroke grestore } bind def
/D7 { gsave translate 45 rotate 0 0 S7 stroke grestore } bind def
/D8 { gsave translate 45 rotate 0 0 S8 stroke grestore } bind def
/D9 { gsave translate 45 rotate 0 0 S9 stroke grestore } bind def
/D10 { gsave translate 45 rotate 0 0 S10 stroke grestore } bind def
/D11 { gsave translate 45 rotate 0 0 S11 stroke grestore } bind def
/D12 { gsave translate 45 rotate 0 0 S12 stroke grestore } bind def
/D13 { gsave translate 45 rotate 0 0 S13 stroke grestore } bind def
/D14 { gsave translate 45 rotate 0 0 S14 stroke grestore } bind def
/D15 { gsave translate 45 rotate 0 0 S15 stroke grestore } bind def
/DiaE { stroke [] 0 setdash vpt add M
  hpt neg vpt neg V hpt vpt neg V
  hpt vpt V hpt neg vpt V closepath stroke } def
/BoxE { stroke [] 0 setdash exch hpt sub exch vpt add M
  0 vpt2 neg V hpt2 0 V 0 vpt2 V
  hpt2 neg 0 V closepath stroke } def
/TriUE { stroke [] 0 setdash vpt 1.12 mul add M
  hpt neg vpt -1.62 mul V
  hpt 2 mul 0 V
  hpt neg vpt 1.62 mul V closepath stroke } def
/TriDE { stroke [] 0 setdash vpt 1.12 mul sub M
  hpt neg vpt 1.62 mul V
  hpt 2 mul 0 V
  hpt neg vpt -1.62 mul V closepath stroke } def
/PentE { stroke [] 0 setdash gsave
  translate 0 hpt M 4 {72 rotate 0 hpt L} repeat
  closepath stroke grestore } def
/CircE { stroke [] 0 setdash 
  hpt 0 360 arc stroke } def
/Opaque { gsave closepath 1 setgray fill grestore 0 setgray closepath } def
/DiaW { stroke [] 0 setdash vpt add M
  hpt neg vpt neg V hpt vpt neg V
  hpt vpt V hpt neg vpt V Opaque stroke } def
/BoxW { stroke [] 0 setdash exch hpt sub exch vpt add M
  0 vpt2 neg V hpt2 0 V 0 vpt2 V
  hpt2 neg 0 V Opaque stroke } def
/TriUW { stroke [] 0 setdash vpt 1.12 mul add M
  hpt neg vpt -1.62 mul V
  hpt 2 mul 0 V
  hpt neg vpt 1.62 mul V Opaque stroke } def
/TriDW { stroke [] 0 setdash vpt 1.12 mul sub M
  hpt neg vpt 1.62 mul V
  hpt 2 mul 0 V
  hpt neg vpt -1.62 mul V Opaque stroke } def
/PentW { stroke [] 0 setdash gsave
  translate 0 hpt M 4 {72 rotate 0 hpt L} repeat
  Opaque stroke grestore } def
/CircW { stroke [] 0 setdash 
  hpt 0 360 arc Opaque stroke } def
/BoxFill { gsave Rec 1 setgray fill grestore } def
/BoxColFill {
  gsave Rec
  /Fillden exch def
  currentrgbcolor
  /ColB exch def /ColG exch def /ColR exch def
  /ColR ColR Fillden mul Fillden sub 1 add def
  /ColG ColG Fillden mul Fillden sub 1 add def
  /ColB ColB Fillden mul Fillden sub 1 add def
  ColR ColG ColB setrgbcolor
  fill grestore } def
%
% PostScript Level 1 Pattern Fill routine
% Usage: x y w h s a XX PatternFill
%	x,y = lower left corner of box to be filled
%	w,h = width and height of box
%	  a = angle in degrees between lines and x-axis
%	 XX = 0/1 for no/yes cross-hatch
%
/PatternFill { gsave /PFa [ 9 2 roll ] def
    PFa 0 get PFa 2 get 2 div add PFa 1 get PFa 3 get 2 div add translate
    PFa 2 get -2 div PFa 3 get -2 div PFa 2 get PFa 3 get Rec
    gsave 1 setgray fill grestore clip
    currentlinewidth 0.5 mul setlinewidth
    /PFs PFa 2 get dup mul PFa 3 get dup mul add sqrt def
    0 0 M PFa 5 get rotate PFs -2 div dup translate
	0 1 PFs PFa 4 get div 1 add floor cvi
	{ PFa 4 get mul 0 M 0 PFs V } for
    0 PFa 6 get ne {
	0 1 PFs PFa 4 get div 1 add floor cvi
	{ PFa 4 get mul 0 2 1 roll M PFs 0 V } for
    } if
    stroke grestore } def
/Symbol-Oblique /Symbol findfont [1 0 .167 1 0 0] makefont
dup length dict begin {1 index /FID eq {pop pop} {def} ifelse} forall
currentdict end definefont pop
end
gnudict begin
gsave
0 0 translate
0.100 0.100 scale
0 setgray
newpath
1.000 UL
LTb
350 200 M
63 0 V
1027 0 R
-63 0 V
1.000 UL
LTb
350 395 M
63 0 V
1027 0 R
-63 0 V
1.000 UL
LTb
350 590 M
63 0 V
1027 0 R
-63 0 V
1.000 UL
LTb
350 785 M
63 0 V
1027 0 R
-63 0 V
1.000 UL
LTb
350 980 M
63 0 V
1027 0 R
-63 0 V
1.000 UL
LTb
350 200 M
0 63 V
0 717 R
0 -63 V
1.000 UL
LTb
568 200 M
0 63 V
0 717 R
0 -63 V
1.000 UL
LTb
786 200 M
0 63 V
0 717 R
0 -63 V
1.000 UL
LTb
1004 200 M
0 63 V
0 717 R
0 -63 V
1.000 UL
LTb
1222 200 M
0 63 V
0 717 R
0 -63 V
1.000 UL
LTb
1440 200 M
0 63 V
0 717 R
0 -63 V
1.000 UL
LTb
1.000 UL
LTa
568 200 M
0 780 V
1.000 UL
LTb
350 200 M
1090 0 V
0 780 V
350 980 L
0 -780 V
0.600 UP
LTb
LTb
LTb
1.000 UL
LT0
1440 980 M
-1 -1 V
-1 -1 V
-1 -1 V
-1 -1 V
-1 -1 V
-2 -1 V
-1 -1 V
-1 -1 V
-1 -1 V
-1 -1 V
-1 -1 V
-1 -1 V
-1 -1 V
-1 -1 V
-2 -1 V
-1 -1 V
-1 -1 V
-1 -1 V
-1 0 V
-1 -1 V
-1 -1 V
-1 -1 V
-2 -1 V
-1 -1 V
-1 -1 V
-1 -1 V
-1 -1 V
-1 -1 V
-1 -1 V
-1 -1 V
-2 -1 V
-1 -1 V
-1 -1 V
-1 -1 V
-1 -1 V
-1 -1 V
-1 -1 V
-2 -1 V
-1 -1 V
-1 -1 V
-1 -1 V
-1 -1 V
-1 -1 V
-2 -1 V
-1 -1 V
-1 -1 V
-1 -1 V
-1 -1 V
-1 -1 V
-2 -1 V
-1 -1 V
-1 -1 V
-1 -1 V
-1 -1 V
-2 -1 V
-1 -1 V
-1 -1 V
-1 -1 V
-1 -1 V
-1 -1 V
-2 -1 V
-1 -1 V
-1 -1 V
-1 -1 V
-1 -1 V
-2 -1 V
-1 -1 V
-1 -1 V
-1 -1 V
-1 -1 V
-2 -1 V
-1 -1 V
-1 -1 V
-1 -1 V
-2 -1 V
-1 -1 V
-1 -1 V
-1 -1 V
-1 -1 V
-2 -1 V
-1 -1 V
-1 -1 V
-1 -1 V
-2 -1 V
-1 -1 V
-1 -1 V
-1 -1 V
-2 -1 V
-1 -1 V
-1 -1 V
-1 -1 V
-2 -1 V
-1 -1 V
-1 -1 V
-1 -1 V
-2 -1 V
-1 -1 V
-1 -1 V
-1 -1 V
-2 -1 V
-1 -1 V
-1 -1 V
-1 -1 V
-2 -1 V
stroke
1317 877 M
-1 -1 V
-1 -1 V
-2 -1 V
-1 -1 V
-1 -2 V
-1 -1 V
-2 -1 V
-1 -1 V
-1 -1 V
-2 -1 V
-1 -1 V
-1 -1 V
-1 -1 V
-2 -1 V
-1 -1 V
-1 -1 V
-2 -1 V
-1 -1 V
-1 -1 V
-2 -1 V
-1 -1 V
-1 -1 V
-2 -1 V
-1 -1 V
-1 -1 V
-2 -1 V
-1 -1 V
-1 -1 V
-2 -1 V
-1 -2 V
-1 -1 V
-2 -1 V
-1 -1 V
-1 -1 V
-2 -1 V
-1 -1 V
-1 -1 V
-2 -1 V
-1 -1 V
-1 -1 V
-2 -1 V
-1 -1 V
-1 -1 V
-2 -1 V
-1 -1 V
-1 -1 V
-2 -1 V
-1 -2 V
-2 -1 V
-1 -1 V
-1 -1 V
-2 -1 V
-1 -1 V
-1 -1 V
-2 -1 V
-1 -1 V
-2 -1 V
-1 -1 V
-1 -1 V
-2 -1 V
-1 -1 V
-2 -1 V
-1 -2 V
-1 -1 V
-2 -1 V
-1 -1 V
-2 -1 V
-1 -1 V
-1 -1 V
-2 -1 V
-1 -1 V
-2 -1 V
-1 -1 V
-2 -1 V
-1 -2 V
-1 -1 V
-2 -1 V
-1 -1 V
-2 -1 V
-1 -1 V
-2 -1 V
-1 -1 V
-2 -1 V
-1 -1 V
-1 -1 V
-2 -1 V
-1 -2 V
-2 -1 V
-1 -1 V
-2 -1 V
-1 -1 V
-2 -1 V
-1 -1 V
-2 -1 V
-1 -1 V
-2 -1 V
-1 -2 V
-2 -1 V
-1 -1 V
-2 -1 V
-1 -1 V
-2 -1 V
-1 -1 V
-2 -1 V
stroke
1173 766 M
-1 -1 V
-2 -1 V
-1 -2 V
-2 -1 V
-1 -1 V
-2 -1 V
-1 -1 V
-2 -1 V
-1 -1 V
-2 -1 V
-1 -1 V
-2 -2 V
-1 -1 V
-2 -1 V
-1 -1 V
-2 -1 V
-2 -1 V
-1 -1 V
-2 -1 V
-1 -2 V
-2 -1 V
-1 -1 V
-2 -1 V
-1 -1 V
-2 -1 V
-2 -1 V
-1 -1 V
-2 -2 V
-1 -1 V
-2 -1 V
-1 -1 V
-2 -1 V
-2 -1 V
-1 -1 V
-2 -2 V
-1 -1 V
-2 -1 V
-2 -1 V
-1 -1 V
-2 -1 V
-1 -1 V
-2 -2 V
-2 -1 V
-1 -1 V
-2 -1 V
-2 -1 V
-1 -1 V
-2 -1 V
-1 -2 V
-2 -1 V
-2 -1 V
-1 -1 V
-2 -1 V
-2 -1 V
-1 -2 V
-2 -1 V
-2 -1 V
-1 -1 V
-2 -1 V
-2 -1 V
-1 -2 V
-2 -1 V
-2 -1 V
-1 -1 V
-2 -1 V
-2 -1 V
-2 -2 V
-1 -1 V
-2 -1 V
-2 -1 V
-1 -1 V
-2 -2 V
-2 -1 V
-2 -1 V
-1 -1 V
-2 -1 V
-2 -1 V
-1 -2 V
-2 -1 V
-2 -1 V
-2 -1 V
-1 -1 V
-2 -2 V
-2 -1 V
-2 -1 V
-1 -1 V
-2 -1 V
-2 -2 V
-2 -1 V
-2 -1 V
-1 -1 V
-2 -1 V
-2 -2 V
-2 -1 V
-2 -1 V
-1 -1 V
-2 -2 V
-2 -1 V
-2 -1 V
-2 -1 V
-2 -1 V
-1 -2 V
-2 -1 V
-2 -1 V
stroke
1001 645 M
-2 -1 V
-2 -2 V
-2 -1 V
-1 -1 V
-2 -1 V
-2 -2 V
-2 -1 V
-2 -1 V
-2 -1 V
-2 -2 V
-2 -1 V
-2 -1 V
-1 -1 V
-2 -2 V
-2 -1 V
-2 -1 V
-2 -1 V
-2 -2 V
-2 -1 V
-2 -1 V
-2 -1 V
-2 -2 V
-2 -1 V
-2 -1 V
-2 -2 V
-2 -1 V
-2 -1 V
-2 -1 V
-2 -2 V
-2 -1 V
-2 -1 V
-2 -2 V
-2 -1 V
-2 -1 V
-2 -2 V
-2 -1 V
-2 -1 V
-2 -2 V
-2 -1 V
-2 -1 V
-2 -2 V
-3 -1 V
-2 -1 V
-2 -2 V
-2 -1 V
-2 -1 V
-2 -2 V
-2 -1 V
-2 -1 V
-3 -2 V
-2 -1 V
-2 -2 V
-2 -1 V
-2 -1 V
-3 -2 V
-2 -1 V
-2 -2 V
-2 -1 V
-3 -1 V
-2 -2 V
-2 -1 V
-2 -2 V
-3 -1 V
-2 -2 V
-2 -1 V
-3 -2 V
-2 -1 V
-2 -1 V
-3 -2 V
-2 -1 V
-3 -2 V
-2 -1 V
-3 -2 V
-2 -1 V
-2 -2 V
-3 -2 V
-3 -1 V
-2 -2 V
-3 -1 V
-2 -2 V
-3 -1 V
-2 -2 V
-3 -2 V
-3 -1 V
-2 -2 V
-3 -2 V
-3 -1 V
-3 -2 V
-2 -2 V
-3 -1 V
-3 -2 V
-3 -2 V
-3 -2 V
-3 -1 V
-3 -2 V
-3 -2 V
-3 -2 V
-4 -2 V
-3 -2 V
-3 -2 V
-3 -2 V
-4 -2 V
-3 -2 V
-4 -2 V
stroke
759 493 M
-4 -3 V
-4 -2 V
-4 -2 V
-4 -3 V
-4 -2 V
-5 -3 V
-4 -3 V
-6 -3 V
-5 -3 V
-6 -4 V
-8 -4 V
-9 -5 V
1.000 UL
LTb
350 200 M
1090 0 V
0 780 V
350 980 L
0 -780 V
0.600 UP
stroke
grestore
end
showpage
}}%
\put(841,824){\makebox(0,0)[l]{(a)}}%
\put(394,863){\makebox(0,0)[l]{$L_1$}}%
\put(1266,278){\makebox(0,0)[l]{$s_{\rm cav}$}}%
\put(1440,100){\makebox(0,0){ 0.2}}%
\put(1222,100){\makebox(0,0){ 0.15}}%
\put(1004,100){\makebox(0,0){ 0.1}}%
\put(786,100){\makebox(0,0){ 0.05}}%
\put(568,100){\makebox(0,0){ 0}}%
\put(350,100){\makebox(0,0){-0.05}}%
\put(300,980){\makebox(0,0)[r]{ 0.8}}%
\put(300,785){\makebox(0,0)[r]{ 0.6}}%
\put(300,590){\makebox(0,0)[r]{ 0.4}}%
\put(300,395){\makebox(0,0)[r]{ 0.2}}%
\put(300,200){\makebox(0,0)[r]{ 0}}%
\end{picture}%
\endgroup
 

%% file: largedevbec2.tex
% GNUPLOT: LaTeX picture with Postscript
\begingroup%
  \makeatletter%
  \newcommand{\GNUPLOTspecial}{%
    \@sanitize\catcode`\%=14\relax\special}%
  \setlength{\unitlength}{0.1bp}%
\begin{picture}(1440,1080)(0,0)%
{\GNUPLOTspecial{"
%!PS-Adobe-2.0 EPSF-2.0
%%Title: largedevbec2.tex
%%Creator: gnuplot 4.0 patchlevel 0
%%CreationDate: Wed May 17 12:13:55 2006
%%DocumentFonts: 
%%BoundingBox: 0 0 144 108
%%Orientation: Landscape
%%EndComments
/gnudict 256 dict def
gnudict begin
/Color true def
/Solid false def
/gnulinewidth 5.000 def
/userlinewidth gnulinewidth def
/vshift -33 def
/dl {10.0 mul} def
/hpt_ 31.5 def
/vpt_ 31.5 def
/hpt hpt_ def
/vpt vpt_ def
/Rounded false def
/M {moveto} bind def
/L {lineto} bind def
/R {rmoveto} bind def
/V {rlineto} bind def
/N {newpath moveto} bind def
/C {setrgbcolor} bind def
/f {rlineto fill} bind def
/vpt2 vpt 2 mul def
/hpt2 hpt 2 mul def
/Lshow { currentpoint stroke M
  0 vshift R show } def
/Rshow { currentpoint stroke M
  dup stringwidth pop neg vshift R show } def
/Cshow { currentpoint stroke M
  dup stringwidth pop -2 div vshift R show } def
/UP { dup vpt_ mul /vpt exch def hpt_ mul /hpt exch def
  /hpt2 hpt 2 mul def /vpt2 vpt 2 mul def } def
/DL { Color {setrgbcolor Solid {pop []} if 0 setdash }
 {pop pop pop 0 setgray Solid {pop []} if 0 setdash} ifelse } def
/BL { stroke userlinewidth 2 mul setlinewidth
      Rounded { 1 setlinejoin 1 setlinecap } if } def
/AL { stroke userlinewidth 2 div setlinewidth
      Rounded { 1 setlinejoin 1 setlinecap } if } def
/UL { dup gnulinewidth mul /userlinewidth exch def
      dup 1 lt {pop 1} if 10 mul /udl exch def } def
/PL { stroke userlinewidth setlinewidth
      Rounded { 1 setlinejoin 1 setlinecap } if } def
/LTw { PL [] 1 setgray } def
/LTb { BL [] 0 0 0 DL } def
/LTa { AL [1 udl mul 2 udl mul] 0 setdash 0 0 0 setrgbcolor } def
/LT0 { PL [] 1 0 0 DL } def
/LT1 { PL [4 dl 2 dl] 0 1 0 DL } def
/LT2 { PL [2 dl 3 dl] 0 0 1 DL } def
/LT3 { PL [1 dl 1.5 dl] 1 0 1 DL } def
/LT4 { PL [5 dl 2 dl 1 dl 2 dl] 0 1 1 DL } def
/LT5 { PL [4 dl 3 dl 1 dl 3 dl] 1 1 0 DL } def
/LT6 { PL [2 dl 2 dl 2 dl 4 dl] 0 0 0 DL } def
/LT7 { PL [2 dl 2 dl 2 dl 2 dl 2 dl 4 dl] 1 0.3 0 DL } def
/LT8 { PL [2 dl 2 dl 2 dl 2 dl 2 dl 2 dl 2 dl 4 dl] 0.5 0.5 0.5 DL } def
/Pnt { stroke [] 0 setdash
   gsave 1 setlinecap M 0 0 V stroke grestore } def
/Dia { stroke [] 0 setdash 2 copy vpt add M
  hpt neg vpt neg V hpt vpt neg V
  hpt vpt V hpt neg vpt V closepath stroke
  Pnt } def
/Pls { stroke [] 0 setdash vpt sub M 0 vpt2 V
  currentpoint stroke M
  hpt neg vpt neg R hpt2 0 V stroke
  } def
/Box { stroke [] 0 setdash 2 copy exch hpt sub exch vpt add M
  0 vpt2 neg V hpt2 0 V 0 vpt2 V
  hpt2 neg 0 V closepath stroke
  Pnt } def
/Crs { stroke [] 0 setdash exch hpt sub exch vpt add M
  hpt2 vpt2 neg V currentpoint stroke M
  hpt2 neg 0 R hpt2 vpt2 V stroke } def
/TriU { stroke [] 0 setdash 2 copy vpt 1.12 mul add M
  hpt neg vpt -1.62 mul V
  hpt 2 mul 0 V
  hpt neg vpt 1.62 mul V closepath stroke
  Pnt  } def
/Star { 2 copy Pls Crs } def
/BoxF { stroke [] 0 setdash exch hpt sub exch vpt add M
  0 vpt2 neg V  hpt2 0 V  0 vpt2 V
  hpt2 neg 0 V  closepath fill } def
/TriUF { stroke [] 0 setdash vpt 1.12 mul add M
  hpt neg vpt -1.62 mul V
  hpt 2 mul 0 V
  hpt neg vpt 1.62 mul V closepath fill } def
/TriD { stroke [] 0 setdash 2 copy vpt 1.12 mul sub M
  hpt neg vpt 1.62 mul V
  hpt 2 mul 0 V
  hpt neg vpt -1.62 mul V closepath stroke
  Pnt  } def
/TriDF { stroke [] 0 setdash vpt 1.12 mul sub M
  hpt neg vpt 1.62 mul V
  hpt 2 mul 0 V
  hpt neg vpt -1.62 mul V closepath fill} def
/DiaF { stroke [] 0 setdash vpt add M
  hpt neg vpt neg V hpt vpt neg V
  hpt vpt V hpt neg vpt V closepath fill } def
/Pent { stroke [] 0 setdash 2 copy gsave
  translate 0 hpt M 4 {72 rotate 0 hpt L} repeat
  closepath stroke grestore Pnt } def
/PentF { stroke [] 0 setdash gsave
  translate 0 hpt M 4 {72 rotate 0 hpt L} repeat
  closepath fill grestore } def
/Circle { stroke [] 0 setdash 2 copy
  hpt 0 360 arc stroke Pnt } def
/CircleF { stroke [] 0 setdash hpt 0 360 arc fill } def
/C0 { BL [] 0 setdash 2 copy moveto vpt 90 450  arc } bind def
/C1 { BL [] 0 setdash 2 copy        moveto
       2 copy  vpt 0 90 arc closepath fill
               vpt 0 360 arc closepath } bind def
/C2 { BL [] 0 setdash 2 copy moveto
       2 copy  vpt 90 180 arc closepath fill
               vpt 0 360 arc closepath } bind def
/C3 { BL [] 0 setdash 2 copy moveto
       2 copy  vpt 0 180 arc closepath fill
               vpt 0 360 arc closepath } bind def
/C4 { BL [] 0 setdash 2 copy moveto
       2 copy  vpt 180 270 arc closepath fill
               vpt 0 360 arc closepath } bind def
/C5 { BL [] 0 setdash 2 copy moveto
       2 copy  vpt 0 90 arc
       2 copy moveto
       2 copy  vpt 180 270 arc closepath fill
               vpt 0 360 arc } bind def
/C6 { BL [] 0 setdash 2 copy moveto
      2 copy  vpt 90 270 arc closepath fill
              vpt 0 360 arc closepath } bind def
/C7 { BL [] 0 setdash 2 copy moveto
      2 copy  vpt 0 270 arc closepath fill
              vpt 0 360 arc closepath } bind def
/C8 { BL [] 0 setdash 2 copy moveto
      2 copy vpt 270 360 arc closepath fill
              vpt 0 360 arc closepath } bind def
/C9 { BL [] 0 setdash 2 copy moveto
      2 copy  vpt 270 450 arc closepath fill
              vpt 0 360 arc closepath } bind def
/C10 { BL [] 0 setdash 2 copy 2 copy moveto vpt 270 360 arc closepath fill
       2 copy moveto
       2 copy vpt 90 180 arc closepath fill
               vpt 0 360 arc closepath } bind def
/C11 { BL [] 0 setdash 2 copy moveto
       2 copy  vpt 0 180 arc closepath fill
       2 copy moveto
       2 copy  vpt 270 360 arc closepath fill
               vpt 0 360 arc closepath } bind def
/C12 { BL [] 0 setdash 2 copy moveto
       2 copy  vpt 180 360 arc closepath fill
               vpt 0 360 arc closepath } bind def
/C13 { BL [] 0 setdash  2 copy moveto
       2 copy  vpt 0 90 arc closepath fill
       2 copy moveto
       2 copy  vpt 180 360 arc closepath fill
               vpt 0 360 arc closepath } bind def
/C14 { BL [] 0 setdash 2 copy moveto
       2 copy  vpt 90 360 arc closepath fill
               vpt 0 360 arc } bind def
/C15 { BL [] 0 setdash 2 copy vpt 0 360 arc closepath fill
               vpt 0 360 arc closepath } bind def
/Rec   { newpath 4 2 roll moveto 1 index 0 rlineto 0 exch rlineto
       neg 0 rlineto closepath } bind def
/Square { dup Rec } bind def
/Bsquare { vpt sub exch vpt sub exch vpt2 Square } bind def
/S0 { BL [] 0 setdash 2 copy moveto 0 vpt rlineto BL Bsquare } bind def
/S1 { BL [] 0 setdash 2 copy vpt Square fill Bsquare } bind def
/S2 { BL [] 0 setdash 2 copy exch vpt sub exch vpt Square fill Bsquare } bind def
/S3 { BL [] 0 setdash 2 copy exch vpt sub exch vpt2 vpt Rec fill Bsquare } bind def
/S4 { BL [] 0 setdash 2 copy exch vpt sub exch vpt sub vpt Square fill Bsquare } bind def
/S5 { BL [] 0 setdash 2 copy 2 copy vpt Square fill
       exch vpt sub exch vpt sub vpt Square fill Bsquare } bind def
/S6 { BL [] 0 setdash 2 copy exch vpt sub exch vpt sub vpt vpt2 Rec fill Bsquare } bind def
/S7 { BL [] 0 setdash 2 copy exch vpt sub exch vpt sub vpt vpt2 Rec fill
       2 copy vpt Square fill
       Bsquare } bind def
/S8 { BL [] 0 setdash 2 copy vpt sub vpt Square fill Bsquare } bind def
/S9 { BL [] 0 setdash 2 copy vpt sub vpt vpt2 Rec fill Bsquare } bind def
/S10 { BL [] 0 setdash 2 copy vpt sub vpt Square fill 2 copy exch vpt sub exch vpt Square fill
       Bsquare } bind def
/S11 { BL [] 0 setdash 2 copy vpt sub vpt Square fill 2 copy exch vpt sub exch vpt2 vpt Rec fill
       Bsquare } bind def
/S12 { BL [] 0 setdash 2 copy exch vpt sub exch vpt sub vpt2 vpt Rec fill Bsquare } bind def
/S13 { BL [] 0 setdash 2 copy exch vpt sub exch vpt sub vpt2 vpt Rec fill
       2 copy vpt Square fill Bsquare } bind def
/S14 { BL [] 0 setdash 2 copy exch vpt sub exch vpt sub vpt2 vpt Rec fill
       2 copy exch vpt sub exch vpt Square fill Bsquare } bind def
/S15 { BL [] 0 setdash 2 copy Bsquare fill Bsquare } bind def
/D0 { gsave translate 45 rotate 0 0 S0 stroke grestore } bind def
/D1 { gsave translate 45 rotate 0 0 S1 stroke grestore } bind def
/D2 { gsave translate 45 rotate 0 0 S2 stroke grestore } bind def
/D3 { gsave translate 45 rotate 0 0 S3 stroke grestore } bind def
/D4 { gsave translate 45 rotate 0 0 S4 stroke grestore } bind def
/D5 { gsave translate 45 rotate 0 0 S5 stroke grestore } bind def
/D6 { gsave translate 45 rotate 0 0 S6 stroke grestore } bind def
/D7 { gsave translate 45 rotate 0 0 S7 stroke grestore } bind def
/D8 { gsave translate 45 rotate 0 0 S8 stroke grestore } bind def
/D9 { gsave translate 45 rotate 0 0 S9 stroke grestore } bind def
/D10 { gsave translate 45 rotate 0 0 S10 stroke grestore } bind def
/D11 { gsave translate 45 rotate 0 0 S11 stroke grestore } bind def
/D12 { gsave translate 45 rotate 0 0 S12 stroke grestore } bind def
/D13 { gsave translate 45 rotate 0 0 S13 stroke grestore } bind def
/D14 { gsave translate 45 rotate 0 0 S14 stroke grestore } bind def
/D15 { gsave translate 45 rotate 0 0 S15 stroke grestore } bind def
/DiaE { stroke [] 0 setdash vpt add M
  hpt neg vpt neg V hpt vpt neg V
  hpt vpt V hpt neg vpt V closepath stroke } def
/BoxE { stroke [] 0 setdash exch hpt sub exch vpt add M
  0 vpt2 neg V hpt2 0 V 0 vpt2 V
  hpt2 neg 0 V closepath stroke } def
/TriUE { stroke [] 0 setdash vpt 1.12 mul add M
  hpt neg vpt -1.62 mul V
  hpt 2 mul 0 V
  hpt neg vpt 1.62 mul V closepath stroke } def
/TriDE { stroke [] 0 setdash vpt 1.12 mul sub M
  hpt neg vpt 1.62 mul V
  hpt 2 mul 0 V
  hpt neg vpt -1.62 mul V closepath stroke } def
/PentE { stroke [] 0 setdash gsave
  translate 0 hpt M 4 {72 rotate 0 hpt L} repeat
  closepath stroke grestore } def
/CircE { stroke [] 0 setdash 
  hpt 0 360 arc stroke } def
/Opaque { gsave closepath 1 setgray fill grestore 0 setgray closepath } def
/DiaW { stroke [] 0 setdash vpt add M
  hpt neg vpt neg V hpt vpt neg V
  hpt vpt V hpt neg vpt V Opaque stroke } def
/BoxW { stroke [] 0 setdash exch hpt sub exch vpt add M
  0 vpt2 neg V hpt2 0 V 0 vpt2 V
  hpt2 neg 0 V Opaque stroke } def
/TriUW { stroke [] 0 setdash vpt 1.12 mul add M
  hpt neg vpt -1.62 mul V
  hpt 2 mul 0 V
  hpt neg vpt 1.62 mul V Opaque stroke } def
/TriDW { stroke [] 0 setdash vpt 1.12 mul sub M
  hpt neg vpt 1.62 mul V
  hpt 2 mul 0 V
  hpt neg vpt -1.62 mul V Opaque stroke } def
/PentW { stroke [] 0 setdash gsave
  translate 0 hpt M 4 {72 rotate 0 hpt L} repeat
  Opaque stroke grestore } def
/CircW { stroke [] 0 setdash 
  hpt 0 360 arc Opaque stroke } def
/BoxFill { gsave Rec 1 setgray fill grestore } def
/BoxColFill {
  gsave Rec
  /Fillden exch def
  currentrgbcolor
  /ColB exch def /ColG exch def /ColR exch def
  /ColR ColR Fillden mul Fillden sub 1 add def
  /ColG ColG Fillden mul Fillden sub 1 add def
  /ColB ColB Fillden mul Fillden sub 1 add def
  ColR ColG ColB setrgbcolor
  fill grestore } def
%
% PostScript Level 1 Pattern Fill routine
% Usage: x y w h s a XX PatternFill
%	x,y = lower left corner of box to be filled
%	w,h = width and height of box
%	  a = angle in degrees between lines and x-axis
%	 XX = 0/1 for no/yes cross-hatch
%
/PatternFill { gsave /PFa [ 9 2 roll ] def
    PFa 0 get PFa 2 get 2 div add PFa 1 get PFa 3 get 2 div add translate
    PFa 2 get -2 div PFa 3 get -2 div PFa 2 get PFa 3 get Rec
    gsave 1 setgray fill grestore clip
    currentlinewidth 0.5 mul setlinewidth
    /PFs PFa 2 get dup mul PFa 3 get dup mul add sqrt def
    0 0 M PFa 5 get rotate PFs -2 div dup translate
	0 1 PFs PFa 4 get div 1 add floor cvi
	{ PFa 4 get mul 0 M 0 PFs V } for
    0 PFa 6 get ne {
	0 1 PFs PFa 4 get div 1 add floor cvi
	{ PFa 4 get mul 0 2 1 roll M PFs 0 V } for
    } if
    stroke grestore } def
/Symbol-Oblique /Symbol findfont [1 0 .167 1 0 0] makefont
dup length dict begin {1 index /FID eq {pop pop} {def} ifelse} forall
currentdict end definefont pop
end
gnudict begin
gsave
0 0 translate
0.100 0.100 scale
0 setgray
newpath
1.000 UL
LTb
400 200 M
63 0 V
977 0 R
-63 0 V
1.000 UL
LTb
400 460 M
63 0 V
977 0 R
-63 0 V
1.000 UL
LTb
400 720 M
63 0 V
977 0 R
-63 0 V
1.000 UL
LTb
400 980 M
63 0 V
977 0 R
-63 0 V
1.000 UL
LTb
400 200 M
0 63 V
0 717 R
0 -63 V
1.000 UL
LTb
608 200 M
0 63 V
0 717 R
0 -63 V
1.000 UL
LTb
816 200 M
0 63 V
0 717 R
0 -63 V
1.000 UL
LTb
1024 200 M
0 63 V
0 717 R
0 -63 V
1.000 UL
LTb
1232 200 M
0 63 V
0 717 R
0 -63 V
1.000 UL
LTb
1440 200 M
0 63 V
0 717 R
0 -63 V
1.000 UL
LTb
1.000 UL
LTa
816 200 M
0 780 V
1.000 UL
LTb
400 200 M
1040 0 V
0 780 V
400 980 L
0 -780 V
0.600 UP
LTb
LTb
LTb
LTb
1.000 UL
LT0
1440 777 M
-3 -2 V
-4 -3 V
-4 -3 V
-4 -3 V
-4 -2 V
-4 -3 V
-4 -3 V
-3 -3 V
-4 -3 V
-4 -3 V
-4 -2 V
-4 -3 V
-4 -3 V
-4 -3 V
-4 -2 V
-4 -3 V
-4 -3 V
-4 -3 V
-4 -3 V
-3 -2 V
-4 -3 V
-4 -3 V
-4 -3 V
-4 -2 V
-4 -3 V
-4 -3 V
-4 -2 V
-4 -3 V
-4 -3 V
-4 -3 V
-4 -2 V
-4 -3 V
-4 -3 V
-4 -2 V
-4 -3 V
-3 -3 V
-4 -2 V
-4 -3 V
-4 -3 V
-4 -2 V
-4 -3 V
-4 -3 V
-4 -2 V
-4 -3 V
-4 -3 V
-4 -2 V
-4 -3 V
-4 -3 V
-4 -2 V
-4 -3 V
-4 -2 V
-4 -3 V
-4 -2 V
-4 -3 V
-4 -3 V
-4 -2 V
-4 -3 V
-4 -2 V
-3 -3 V
-4 -2 V
-4 -3 V
-4 -3 V
-4 -2 V
-4 -3 V
-4 -2 V
-4 -3 V
-4 -2 V
-4 -3 V
-4 -2 V
-4 -3 V
-4 -2 V
-4 -3 V
-4 -2 V
-4 -3 V
-4 -2 V
-4 -3 V
-4 -2 V
-4 -2 V
-4 -3 V
-4 -2 V
-4 -3 V
-4 -2 V
-4 -3 V
-4 -2 V
-4 -2 V
-4 -3 V
-4 -2 V
-4 -3 V
-4 -2 V
-4 -2 V
-4 -3 V
-4 -2 V
-4 -3 V
-4 -2 V
-4 -2 V
-4 -3 V
-4 -2 V
-4 -2 V
-4 -3 V
-4 -2 V
-4 -2 V
-4 -3 V
-4 -2 V
-4 -2 V
stroke
1029 509 M
-4 -3 V
-4 -2 V
-4 -2 V
-4 -2 V
-4 -3 V
-4 -2 V
-4 -2 V
-4 -2 V
-4 -3 V
-4 -2 V
-4 -2 V
-4 -2 V
-4 -3 V
-4 -2 V
-4 -2 V
-4 -2 V
-4 -3 V
-4 -2 V
-4 -2 V
-4 -2 V
-4 -2 V
-4 -3 V
-4 -2 V
-4 -2 V
-4 -2 V
-4 -2 V
-4 -2 V
-4 -3 V
-4 -2 V
-4 -2 V
-4 -2 V
-4 -2 V
-4 -2 V
-4 -2 V
-4 -2 V
-4 -3 V
-4 -2 V
-4 -2 V
-4 -2 V
-4 -2 V
-4 -2 V
-5 -2 V
-4 -2 V
-4 -2 V
-4 -2 V
-4 -2 V
-4 -2 V
-4 -3 V
-4 -2 V
-4 -2 V
-4 -2 V
-4 -2 V
-4 -2 V
-5 -2 V
-4 -2 V
-4 -2 V
-4 -2 V
-4 -2 V
-4 -2 V
-4 -2 V
-5 -2 V
-4 -2 V
-4 -2 V
-4 -2 V
-4 -2 V
-5 -2 V
-4 -2 V
-4 -2 V
-4 -2 V
-5 -2 V
-4 -2 V
-4 -2 V
-5 -2 V
-4 -2 V
-5 -3 V
-4 -2 V
-5 -2 V
-4 -2 V
-5 -2 V
-4 -2 V
-5 -2 V
-5 -2 V
-5 -2 V
-5 -3 V
-5 -2 V
-5 -2 V
-5 -3 V
-6 -2 V
-5 -3 V
-6 -2 V
-7 -3 V
-7 -3 V
-8 -4 V
-11 -4 V
0.600 UP
1.000 UL
LT6
816 394 CircleF
1.000 UL
LTb
400 200 M
1040 0 V
0 780 V
400 980 L
0 -780 V
0.600 UP
stroke
grestore
end
showpage
}}%
\put(629,460){\makebox(0,0)[l]{$E_{\rm av}$}}%
\put(868,824){\makebox(0,0)[l]{(b)}}%
\put(452,824){\makebox(0,0)[l]{$L_1$}}%
\put(1253,304){\makebox(0,0)[l]{$s_{\rm cav}$}}%
\put(1440,100){\makebox(0,0){ 0.06}}%
\put(1232,100){\makebox(0,0){ 0.04}}%
\put(1024,100){\makebox(0,0){ 0.02}}%
\put(816,100){\makebox(0,0){ 0}}%
\put(608,100){\makebox(0,0){-0.02}}%
\put(400,100){\makebox(0,0){-0.04}}%
\put(350,980){\makebox(0,0)[r]{ 0.15}}%
\put(350,720){\makebox(0,0)[r]{ 0.1}}%
\put(350,460){\makebox(0,0)[r]{ 0.05}}%
\put(350,200){\makebox(0,0)[r]{ 0}}%
\end{picture}%
\endgroup
 

%% file: largedevbec3.tex
% GNUPLOT: LaTeX picture with Postscript
\begingroup%
  \makeatletter%
  \newcommand{\GNUPLOTspecial}{%
    \@sanitize\catcode`\%=14\relax\special}%
  \setlength{\unitlength}{0.1bp}%
\begin{picture}(1440,1080)(0,0)%
{\GNUPLOTspecial{"
%!PS-Adobe-2.0 EPSF-2.0
%%Title: largedevbec3.tex
%%Creator: gnuplot 4.0 patchlevel 0
%%CreationDate: Wed May 17 12:13:55 2006
%%DocumentFonts: 
%%BoundingBox: 0 0 144 108
%%Orientation: Landscape
%%EndComments
/gnudict 256 dict def
gnudict begin
/Color true def
/Solid false def
/gnulinewidth 5.000 def
/userlinewidth gnulinewidth def
/vshift -33 def
/dl {10.0 mul} def
/hpt_ 31.5 def
/vpt_ 31.5 def
/hpt hpt_ def
/vpt vpt_ def
/Rounded false def
/M {moveto} bind def
/L {lineto} bind def
/R {rmoveto} bind def
/V {rlineto} bind def
/N {newpath moveto} bind def
/C {setrgbcolor} bind def
/f {rlineto fill} bind def
/vpt2 vpt 2 mul def
/hpt2 hpt 2 mul def
/Lshow { currentpoint stroke M
  0 vshift R show } def
/Rshow { currentpoint stroke M
  dup stringwidth pop neg vshift R show } def
/Cshow { currentpoint stroke M
  dup stringwidth pop -2 div vshift R show } def
/UP { dup vpt_ mul /vpt exch def hpt_ mul /hpt exch def
  /hpt2 hpt 2 mul def /vpt2 vpt 2 mul def } def
/DL { Color {setrgbcolor Solid {pop []} if 0 setdash }
 {pop pop pop 0 setgray Solid {pop []} if 0 setdash} ifelse } def
/BL { stroke userlinewidth 2 mul setlinewidth
      Rounded { 1 setlinejoin 1 setlinecap } if } def
/AL { stroke userlinewidth 2 div setlinewidth
      Rounded { 1 setlinejoin 1 setlinecap } if } def
/UL { dup gnulinewidth mul /userlinewidth exch def
      dup 1 lt {pop 1} if 10 mul /udl exch def } def
/PL { stroke userlinewidth setlinewidth
      Rounded { 1 setlinejoin 1 setlinecap } if } def
/LTw { PL [] 1 setgray } def
/LTb { BL [] 0 0 0 DL } def
/LTa { AL [1 udl mul 2 udl mul] 0 setdash 0 0 0 setrgbcolor } def
/LT0 { PL [] 1 0 0 DL } def
/LT1 { PL [4 dl 2 dl] 0 1 0 DL } def
/LT2 { PL [2 dl 3 dl] 0 0 1 DL } def
/LT3 { PL [1 dl 1.5 dl] 1 0 1 DL } def
/LT4 { PL [5 dl 2 dl 1 dl 2 dl] 0 1 1 DL } def
/LT5 { PL [4 dl 3 dl 1 dl 3 dl] 1 1 0 DL } def
/LT6 { PL [2 dl 2 dl 2 dl 4 dl] 0 0 0 DL } def
/LT7 { PL [2 dl 2 dl 2 dl 2 dl 2 dl 4 dl] 1 0.3 0 DL } def
/LT8 { PL [2 dl 2 dl 2 dl 2 dl 2 dl 2 dl 2 dl 4 dl] 0.5 0.5 0.5 DL } def
/Pnt { stroke [] 0 setdash
   gsave 1 setlinecap M 0 0 V stroke grestore } def
/Dia { stroke [] 0 setdash 2 copy vpt add M
  hpt neg vpt neg V hpt vpt neg V
  hpt vpt V hpt neg vpt V closepath stroke
  Pnt } def
/Pls { stroke [] 0 setdash vpt sub M 0 vpt2 V
  currentpoint stroke M
  hpt neg vpt neg R hpt2 0 V stroke
  } def
/Box { stroke [] 0 setdash 2 copy exch hpt sub exch vpt add M
  0 vpt2 neg V hpt2 0 V 0 vpt2 V
  hpt2 neg 0 V closepath stroke
  Pnt } def
/Crs { stroke [] 0 setdash exch hpt sub exch vpt add M
  hpt2 vpt2 neg V currentpoint stroke M
  hpt2 neg 0 R hpt2 vpt2 V stroke } def
/TriU { stroke [] 0 setdash 2 copy vpt 1.12 mul add M
  hpt neg vpt -1.62 mul V
  hpt 2 mul 0 V
  hpt neg vpt 1.62 mul V closepath stroke
  Pnt  } def
/Star { 2 copy Pls Crs } def
/BoxF { stroke [] 0 setdash exch hpt sub exch vpt add M
  0 vpt2 neg V  hpt2 0 V  0 vpt2 V
  hpt2 neg 0 V  closepath fill } def
/TriUF { stroke [] 0 setdash vpt 1.12 mul add M
  hpt neg vpt -1.62 mul V
  hpt 2 mul 0 V
  hpt neg vpt 1.62 mul V closepath fill } def
/TriD { stroke [] 0 setdash 2 copy vpt 1.12 mul sub M
  hpt neg vpt 1.62 mul V
  hpt 2 mul 0 V
  hpt neg vpt -1.62 mul V closepath stroke
  Pnt  } def
/TriDF { stroke [] 0 setdash vpt 1.12 mul sub M
  hpt neg vpt 1.62 mul V
  hpt 2 mul 0 V
  hpt neg vpt -1.62 mul V closepath fill} def
/DiaF { stroke [] 0 setdash vpt add M
  hpt neg vpt neg V hpt vpt neg V
  hpt vpt V hpt neg vpt V closepath fill } def
/Pent { stroke [] 0 setdash 2 copy gsave
  translate 0 hpt M 4 {72 rotate 0 hpt L} repeat
  closepath stroke grestore Pnt } def
/PentF { stroke [] 0 setdash gsave
  translate 0 hpt M 4 {72 rotate 0 hpt L} repeat
  closepath fill grestore } def
/Circle { stroke [] 0 setdash 2 copy
  hpt 0 360 arc stroke Pnt } def
/CircleF { stroke [] 0 setdash hpt 0 360 arc fill } def
/C0 { BL [] 0 setdash 2 copy moveto vpt 90 450  arc } bind def
/C1 { BL [] 0 setdash 2 copy        moveto
       2 copy  vpt 0 90 arc closepath fill
               vpt 0 360 arc closepath } bind def
/C2 { BL [] 0 setdash 2 copy moveto
       2 copy  vpt 90 180 arc closepath fill
               vpt 0 360 arc closepath } bind def
/C3 { BL [] 0 setdash 2 copy moveto
       2 copy  vpt 0 180 arc closepath fill
               vpt 0 360 arc closepath } bind def
/C4 { BL [] 0 setdash 2 copy moveto
       2 copy  vpt 180 270 arc closepath fill
               vpt 0 360 arc closepath } bind def
/C5 { BL [] 0 setdash 2 copy moveto
       2 copy  vpt 0 90 arc
       2 copy moveto
       2 copy  vpt 180 270 arc closepath fill
               vpt 0 360 arc } bind def
/C6 { BL [] 0 setdash 2 copy moveto
      2 copy  vpt 90 270 arc closepath fill
              vpt 0 360 arc closepath } bind def
/C7 { BL [] 0 setdash 2 copy moveto
      2 copy  vpt 0 270 arc closepath fill
              vpt 0 360 arc closepath } bind def
/C8 { BL [] 0 setdash 2 copy moveto
      2 copy vpt 270 360 arc closepath fill
              vpt 0 360 arc closepath } bind def
/C9 { BL [] 0 setdash 2 copy moveto
      2 copy  vpt 270 450 arc closepath fill
              vpt 0 360 arc closepath } bind def
/C10 { BL [] 0 setdash 2 copy 2 copy moveto vpt 270 360 arc closepath fill
       2 copy moveto
       2 copy vpt 90 180 arc closepath fill
               vpt 0 360 arc closepath } bind def
/C11 { BL [] 0 setdash 2 copy moveto
       2 copy  vpt 0 180 arc closepath fill
       2 copy moveto
       2 copy  vpt 270 360 arc closepath fill
               vpt 0 360 arc closepath } bind def
/C12 { BL [] 0 setdash 2 copy moveto
       2 copy  vpt 180 360 arc closepath fill
               vpt 0 360 arc closepath } bind def
/C13 { BL [] 0 setdash  2 copy moveto
       2 copy  vpt 0 90 arc closepath fill
       2 copy moveto
       2 copy  vpt 180 360 arc closepath fill
               vpt 0 360 arc closepath } bind def
/C14 { BL [] 0 setdash 2 copy moveto
       2 copy  vpt 90 360 arc closepath fill
               vpt 0 360 arc } bind def
/C15 { BL [] 0 setdash 2 copy vpt 0 360 arc closepath fill
               vpt 0 360 arc closepath } bind def
/Rec   { newpath 4 2 roll moveto 1 index 0 rlineto 0 exch rlineto
       neg 0 rlineto closepath } bind def
/Square { dup Rec } bind def
/Bsquare { vpt sub exch vpt sub exch vpt2 Square } bind def
/S0 { BL [] 0 setdash 2 copy moveto 0 vpt rlineto BL Bsquare } bind def
/S1 { BL [] 0 setdash 2 copy vpt Square fill Bsquare } bind def
/S2 { BL [] 0 setdash 2 copy exch vpt sub exch vpt Square fill Bsquare } bind def
/S3 { BL [] 0 setdash 2 copy exch vpt sub exch vpt2 vpt Rec fill Bsquare } bind def
/S4 { BL [] 0 setdash 2 copy exch vpt sub exch vpt sub vpt Square fill Bsquare } bind def
/S5 { BL [] 0 setdash 2 copy 2 copy vpt Square fill
       exch vpt sub exch vpt sub vpt Square fill Bsquare } bind def
/S6 { BL [] 0 setdash 2 copy exch vpt sub exch vpt sub vpt vpt2 Rec fill Bsquare } bind def
/S7 { BL [] 0 setdash 2 copy exch vpt sub exch vpt sub vpt vpt2 Rec fill
       2 copy vpt Square fill
       Bsquare } bind def
/S8 { BL [] 0 setdash 2 copy vpt sub vpt Square fill Bsquare } bind def
/S9 { BL [] 0 setdash 2 copy vpt sub vpt vpt2 Rec fill Bsquare } bind def
/S10 { BL [] 0 setdash 2 copy vpt sub vpt Square fill 2 copy exch vpt sub exch vpt Square fill
       Bsquare } bind def
/S11 { BL [] 0 setdash 2 copy vpt sub vpt Square fill 2 copy exch vpt sub exch vpt2 vpt Rec fill
       Bsquare } bind def
/S12 { BL [] 0 setdash 2 copy exch vpt sub exch vpt sub vpt2 vpt Rec fill Bsquare } bind def
/S13 { BL [] 0 setdash 2 copy exch vpt sub exch vpt sub vpt2 vpt Rec fill
       2 copy vpt Square fill Bsquare } bind def
/S14 { BL [] 0 setdash 2 copy exch vpt sub exch vpt sub vpt2 vpt Rec fill
       2 copy exch vpt sub exch vpt Square fill Bsquare } bind def
/S15 { BL [] 0 setdash 2 copy Bsquare fill Bsquare } bind def
/D0 { gsave translate 45 rotate 0 0 S0 stroke grestore } bind def
/D1 { gsave translate 45 rotate 0 0 S1 stroke grestore } bind def
/D2 { gsave translate 45 rotate 0 0 S2 stroke grestore } bind def
/D3 { gsave translate 45 rotate 0 0 S3 stroke grestore } bind def
/D4 { gsave translate 45 rotate 0 0 S4 stroke grestore } bind def
/D5 { gsave translate 45 rotate 0 0 S5 stroke grestore } bind def
/D6 { gsave translate 45 rotate 0 0 S6 stroke grestore } bind def
/D7 { gsave translate 45 rotate 0 0 S7 stroke grestore } bind def
/D8 { gsave translate 45 rotate 0 0 S8 stroke grestore } bind def
/D9 { gsave translate 45 rotate 0 0 S9 stroke grestore } bind def
/D10 { gsave translate 45 rotate 0 0 S10 stroke grestore } bind def
/D11 { gsave translate 45 rotate 0 0 S11 stroke grestore } bind def
/D12 { gsave translate 45 rotate 0 0 S12 stroke grestore } bind def
/D13 { gsave translate 45 rotate 0 0 S13 stroke grestore } bind def
/D14 { gsave translate 45 rotate 0 0 S14 stroke grestore } bind def
/D15 { gsave translate 45 rotate 0 0 S15 stroke grestore } bind def
/DiaE { stroke [] 0 setdash vpt add M
  hpt neg vpt neg V hpt vpt neg V
  hpt vpt V hpt neg vpt V closepath stroke } def
/BoxE { stroke [] 0 setdash exch hpt sub exch vpt add M
  0 vpt2 neg V hpt2 0 V 0 vpt2 V
  hpt2 neg 0 V closepath stroke } def
/TriUE { stroke [] 0 setdash vpt 1.12 mul add M
  hpt neg vpt -1.62 mul V
  hpt 2 mul 0 V
  hpt neg vpt 1.62 mul V closepath stroke } def
/TriDE { stroke [] 0 setdash vpt 1.12 mul sub M
  hpt neg vpt 1.62 mul V
  hpt 2 mul 0 V
  hpt neg vpt -1.62 mul V closepath stroke } def
/PentE { stroke [] 0 setdash gsave
  translate 0 hpt M 4 {72 rotate 0 hpt L} repeat
  closepath stroke grestore } def
/CircE { stroke [] 0 setdash 
  hpt 0 360 arc stroke } def
/Opaque { gsave closepath 1 setgray fill grestore 0 setgray closepath } def
/DiaW { stroke [] 0 setdash vpt add M
  hpt neg vpt neg V hpt vpt neg V
  hpt vpt V hpt neg vpt V Opaque stroke } def
/BoxW { stroke [] 0 setdash exch hpt sub exch vpt add M
  0 vpt2 neg V hpt2 0 V 0 vpt2 V
  hpt2 neg 0 V Opaque stroke } def
/TriUW { stroke [] 0 setdash vpt 1.12 mul add M
  hpt neg vpt -1.62 mul V
  hpt 2 mul 0 V
  hpt neg vpt 1.62 mul V Opaque stroke } def
/TriDW { stroke [] 0 setdash vpt 1.12 mul sub M
  hpt neg vpt 1.62 mul V
  hpt 2 mul 0 V
  hpt neg vpt -1.62 mul V Opaque stroke } def
/PentW { stroke [] 0 setdash gsave
  translate 0 hpt M 4 {72 rotate 0 hpt L} repeat
  Opaque stroke grestore } def
/CircW { stroke [] 0 setdash 
  hpt 0 360 arc Opaque stroke } def
/BoxFill { gsave Rec 1 setgray fill grestore } def
/BoxColFill {
  gsave Rec
  /Fillden exch def
  currentrgbcolor
  /ColB exch def /ColG exch def /ColR exch def
  /ColR ColR Fillden mul Fillden sub 1 add def
  /ColG ColG Fillden mul Fillden sub 1 add def
  /ColB ColB Fillden mul Fillden sub 1 add def
  ColR ColG ColB setrgbcolor
  fill grestore } def
%
% PostScript Level 1 Pattern Fill routine
% Usage: x y w h s a XX PatternFill
%	x,y = lower left corner of box to be filled
%	w,h = width and height of box
%	  a = angle in degrees between lines and x-axis
%	 XX = 0/1 for no/yes cross-hatch
%
/PatternFill { gsave /PFa [ 9 2 roll ] def
    PFa 0 get PFa 2 get 2 div add PFa 1 get PFa 3 get 2 div add translate
    PFa 2 get -2 div PFa 3 get -2 div PFa 2 get PFa 3 get Rec
    gsave 1 setgray fill grestore clip
    currentlinewidth 0.5 mul setlinewidth
    /PFs PFa 2 get dup mul PFa 3 get dup mul add sqrt def
    0 0 M PFa 5 get rotate PFs -2 div dup translate
	0 1 PFs PFa 4 get div 1 add floor cvi
	{ PFa 4 get mul 0 M 0 PFs V } for
    0 PFa 6 get ne {
	0 1 PFs PFa 4 get div 1 add floor cvi
	{ PFa 4 get mul 0 2 1 roll M PFs 0 V } for
    } if
    stroke grestore } def
/Symbol-Oblique /Symbol findfont [1 0 .167 1 0 0] makefont
dup length dict begin {1 index /FID eq {pop pop} {def} ifelse} forall
currentdict end definefont pop
end
gnudict begin
gsave
0 0 translate
0.100 0.100 scale
0 setgray
newpath
1.000 UL
LTb
450 200 M
63 0 V
927 0 R
-63 0 V
1.000 UL
LTb
450 395 M
63 0 V
927 0 R
-63 0 V
1.000 UL
LTb
450 590 M
63 0 V
927 0 R
-63 0 V
1.000 UL
LTb
450 785 M
63 0 V
927 0 R
-63 0 V
1.000 UL
LTb
450 980 M
63 0 V
927 0 R
-63 0 V
1.000 UL
LTb
574 200 M
0 63 V
0 717 R
0 -63 V
1.000 UL
LTb
821 200 M
0 63 V
0 717 R
0 -63 V
1.000 UL
LTb
1069 200 M
0 63 V
0 717 R
0 -63 V
1.000 UL
LTb
1316 200 M
0 63 V
0 717 R
0 -63 V
1.000 UL
LTb
1.000 UL
LTa
1316 200 M
0 780 V
1.000 UL
LTb
450 200 M
990 0 V
0 780 V
-990 0 V
0 -780 V
0.600 UP
LTb
LTb
LTb
LTb
LTb
1.000 UL
LT0
1423 973 M
-7 -17 V
-7 -17 V
-8 -16 V
-7 -17 V
-7 -16 V
-7 -16 V
-7 -15 V
-7 -16 V
-7 -15 V
-7 -15 V
-7 -15 V
-7 -14 V
-7 -15 V
-7 -14 V
-7 -14 V
-7 -14 V
-7 -14 V
-7 -13 V
-7 -13 V
-7 -13 V
-6 -13 V
-7 -13 V
-7 -12 V
-7 -13 V
-7 -12 V
-6 -11 V
-7 -12 V
-7 -12 V
-6 -11 V
-7 -11 V
-7 -11 V
-6 -11 V
-7 -10 V
-6 -10 V
-7 -11 V
-7 -10 V
-6 -9 V
-7 -10 V
-6 -9 V
-6 -10 V
-7 -9 V
-6 -8 V
-7 -9 V
-6 -9 V
-6 -8 V
-7 -8 V
-6 -8 V
-6 -8 V
-7 -7 V
-6 -8 V
-6 -7 V
-6 -7 V
-6 -7 V
-6 -7 V
-7 -6 V
-6 -7 V
-6 -6 V
-6 -6 V
-6 -6 V
-6 -6 V
-6 -5 V
-6 -6 V
-5 -5 V
-6 -5 V
-6 -5 V
-6 -5 V
-6 -4 V
-6 -5 V
-5 -4 V
-6 -4 V
-6 -4 V
-6 -4 V
-5 -4 V
-6 -3 V
-5 -4 V
-6 -3 V
-5 -3 V
-6 -3 V
-5 -3 V
-6 -2 V
-5 -3 V
-6 -2 V
-5 -2 V
-5 -2 V
-6 -2 V
-5 -2 V
-5 -2 V
-6 -1 V
-5 -2 V
-5 -1 V
-5 -1 V
-5 -1 V
-5 -1 V
-5 -1 V
-5 -1 V
-5 0 V
-5 0 V
-5 -1 V
-5 0 V
-5 0 V
-5 0 V
-4 0 V
-5 1 V
-5 0 V
stroke
789 201 M
-5 0 V
-4 1 V
-5 1 V
-4 1 V
-5 0 V
-4 1 V
-5 2 V
-4 1 V
-5 1 V
-4 1 V
-4 2 V
-5 2 V
-4 1 V
-4 2 V
-4 2 V
-5 2 V
-4 2 V
-4 2 V
-4 2 V
-4 2 V
-4 2 V
-4 3 V
-4 2 V
-3 3 V
-4 2 V
-4 3 V
-4 2 V
-3 3 V
-4 3 V
-3 3 V
-4 3 V
-3 3 V
-4 3 V
-3 3 V
-4 3 V
-3 3 V
-3 3 V
-3 3 V
-4 3 V
-3 3 V
-3 4 V
-3 3 V
-3 3 V
-3 3 V
-2 4 V
-3 3 V
-3 3 V
-3 3 V
-2 4 V
-3 3 V
-2 3 V
-3 3 V
-2 3 V
-2 4 V
-3 3 V
-2 3 V
-2 3 V
-2 3 V
-2 2 V
-2 3 V
-1 3 V
-2 2 V
-2 3 V
-1 2 V
-1 2 V
-2 2 V
-1 2 V
0 1 V
-1 2 V
-1 0 V
0 1 V
0 -1 V
1 -1 V
2 -4 V
0.600 UP
1.000 UL
LT6
1317 741 CircleF
0.600 UP
1.000 UL
LT6
813 200 CircleF
1.000 UL
LTb
450 200 M
990 0 V
0 780 V
-990 0 V
0 -780 V
0.600 UP
stroke
grestore
end
showpage
}}%
\put(747,317){\makebox(0,0)[l]{$\bar s_{\rm cav}$}}%
\put(1094,785){\makebox(0,0)[l]{$E_{\rm av}$}}%
\put(896,824){\makebox(0,0)[l]{(c)}}%
\put(475,883){\makebox(0,0)[l]{$L_1$}}%
\put(1118,259){\makebox(0,0)[l]{$s_{\rm cav}$}}%
\put(1316,100){\makebox(0,0){ 0}}%
\put(1069,100){\makebox(0,0){-0.01}}%
\put(821,100){\makebox(0,0){-0.02}}%
\put(574,100){\makebox(0,0){-0.03}}%
\put(400,980){\makebox(0,0)[r]{ 0.004}}%
\put(400,785){\makebox(0,0)[r]{ 0.003}}%
\put(400,590){\makebox(0,0)[r]{ 0.002}}%
\put(400,395){\makebox(0,0)[r]{ 0.001}}%
\put(400,200){\makebox(0,0)[r]{ 0}}%
\end{picture}%
\endgroup
 

%% file: largedevbec4.tex
% GNUPLOT: LaTeX picture with Postscript
\begingroup%
  \makeatletter%
  \newcommand{\GNUPLOTspecial}{%
    \@sanitize\catcode`\%=14\relax\special}%
  \setlength{\unitlength}{0.1bp}%
\begin{picture}(1440,1080)(0,0)%
{\GNUPLOTspecial{"
%!PS-Adobe-2.0 EPSF-2.0
%%Title: largedevbec4.tex
%%Creator: gnuplot 4.0 patchlevel 0
%%CreationDate: Wed May 17 12:13:55 2006
%%DocumentFonts: 
%%BoundingBox: 0 0 144 108
%%Orientation: Landscape
%%EndComments
/gnudict 256 dict def
gnudict begin
/Color true def
/Solid false def
/gnulinewidth 5.000 def
/userlinewidth gnulinewidth def
/vshift -33 def
/dl {10.0 mul} def
/hpt_ 31.5 def
/vpt_ 31.5 def
/hpt hpt_ def
/vpt vpt_ def
/Rounded false def
/M {moveto} bind def
/L {lineto} bind def
/R {rmoveto} bind def
/V {rlineto} bind def
/N {newpath moveto} bind def
/C {setrgbcolor} bind def
/f {rlineto fill} bind def
/vpt2 vpt 2 mul def
/hpt2 hpt 2 mul def
/Lshow { currentpoint stroke M
  0 vshift R show } def
/Rshow { currentpoint stroke M
  dup stringwidth pop neg vshift R show } def
/Cshow { currentpoint stroke M
  dup stringwidth pop -2 div vshift R show } def
/UP { dup vpt_ mul /vpt exch def hpt_ mul /hpt exch def
  /hpt2 hpt 2 mul def /vpt2 vpt 2 mul def } def
/DL { Color {setrgbcolor Solid {pop []} if 0 setdash }
 {pop pop pop 0 setgray Solid {pop []} if 0 setdash} ifelse } def
/BL { stroke userlinewidth 2 mul setlinewidth
      Rounded { 1 setlinejoin 1 setlinecap } if } def
/AL { stroke userlinewidth 2 div setlinewidth
      Rounded { 1 setlinejoin 1 setlinecap } if } def
/UL { dup gnulinewidth mul /userlinewidth exch def
      dup 1 lt {pop 1} if 10 mul /udl exch def } def
/PL { stroke userlinewidth setlinewidth
      Rounded { 1 setlinejoin 1 setlinecap } if } def
/LTw { PL [] 1 setgray } def
/LTb { BL [] 0 0 0 DL } def
/LTa { AL [1 udl mul 2 udl mul] 0 setdash 0 0 0 setrgbcolor } def
/LT0 { PL [] 1 0 0 DL } def
/LT1 { PL [4 dl 2 dl] 0 1 0 DL } def
/LT2 { PL [2 dl 3 dl] 0 0 1 DL } def
/LT3 { PL [1 dl 1.5 dl] 1 0 1 DL } def
/LT4 { PL [5 dl 2 dl 1 dl 2 dl] 0 1 1 DL } def
/LT5 { PL [4 dl 3 dl 1 dl 3 dl] 1 1 0 DL } def
/LT6 { PL [2 dl 2 dl 2 dl 4 dl] 0 0 0 DL } def
/LT7 { PL [2 dl 2 dl 2 dl 2 dl 2 dl 4 dl] 1 0.3 0 DL } def
/LT8 { PL [2 dl 2 dl 2 dl 2 dl 2 dl 2 dl 2 dl 4 dl] 0.5 0.5 0.5 DL } def
/Pnt { stroke [] 0 setdash
   gsave 1 setlinecap M 0 0 V stroke grestore } def
/Dia { stroke [] 0 setdash 2 copy vpt add M
  hpt neg vpt neg V hpt vpt neg V
  hpt vpt V hpt neg vpt V closepath stroke
  Pnt } def
/Pls { stroke [] 0 setdash vpt sub M 0 vpt2 V
  currentpoint stroke M
  hpt neg vpt neg R hpt2 0 V stroke
  } def
/Box { stroke [] 0 setdash 2 copy exch hpt sub exch vpt add M
  0 vpt2 neg V hpt2 0 V 0 vpt2 V
  hpt2 neg 0 V closepath stroke
  Pnt } def
/Crs { stroke [] 0 setdash exch hpt sub exch vpt add M
  hpt2 vpt2 neg V currentpoint stroke M
  hpt2 neg 0 R hpt2 vpt2 V stroke } def
/TriU { stroke [] 0 setdash 2 copy vpt 1.12 mul add M
  hpt neg vpt -1.62 mul V
  hpt 2 mul 0 V
  hpt neg vpt 1.62 mul V closepath stroke
  Pnt  } def
/Star { 2 copy Pls Crs } def
/BoxF { stroke [] 0 setdash exch hpt sub exch vpt add M
  0 vpt2 neg V  hpt2 0 V  0 vpt2 V
  hpt2 neg 0 V  closepath fill } def
/TriUF { stroke [] 0 setdash vpt 1.12 mul add M
  hpt neg vpt -1.62 mul V
  hpt 2 mul 0 V
  hpt neg vpt 1.62 mul V closepath fill } def
/TriD { stroke [] 0 setdash 2 copy vpt 1.12 mul sub M
  hpt neg vpt 1.62 mul V
  hpt 2 mul 0 V
  hpt neg vpt -1.62 mul V closepath stroke
  Pnt  } def
/TriDF { stroke [] 0 setdash vpt 1.12 mul sub M
  hpt neg vpt 1.62 mul V
  hpt 2 mul 0 V
  hpt neg vpt -1.62 mul V closepath fill} def
/DiaF { stroke [] 0 setdash vpt add M
  hpt neg vpt neg V hpt vpt neg V
  hpt vpt V hpt neg vpt V closepath fill } def
/Pent { stroke [] 0 setdash 2 copy gsave
  translate 0 hpt M 4 {72 rotate 0 hpt L} repeat
  closepath stroke grestore Pnt } def
/PentF { stroke [] 0 setdash gsave
  translate 0 hpt M 4 {72 rotate 0 hpt L} repeat
  closepath fill grestore } def
/Circle { stroke [] 0 setdash 2 copy
  hpt 0 360 arc stroke Pnt } def
/CircleF { stroke [] 0 setdash hpt 0 360 arc fill } def
/C0 { BL [] 0 setdash 2 copy moveto vpt 90 450  arc } bind def
/C1 { BL [] 0 setdash 2 copy        moveto
       2 copy  vpt 0 90 arc closepath fill
               vpt 0 360 arc closepath } bind def
/C2 { BL [] 0 setdash 2 copy moveto
       2 copy  vpt 90 180 arc closepath fill
               vpt 0 360 arc closepath } bind def
/C3 { BL [] 0 setdash 2 copy moveto
       2 copy  vpt 0 180 arc closepath fill
               vpt 0 360 arc closepath } bind def
/C4 { BL [] 0 setdash 2 copy moveto
       2 copy  vpt 180 270 arc closepath fill
               vpt 0 360 arc closepath } bind def
/C5 { BL [] 0 setdash 2 copy moveto
       2 copy  vpt 0 90 arc
       2 copy moveto
       2 copy  vpt 180 270 arc closepath fill
               vpt 0 360 arc } bind def
/C6 { BL [] 0 setdash 2 copy moveto
      2 copy  vpt 90 270 arc closepath fill
              vpt 0 360 arc closepath } bind def
/C7 { BL [] 0 setdash 2 copy moveto
      2 copy  vpt 0 270 arc closepath fill
              vpt 0 360 arc closepath } bind def
/C8 { BL [] 0 setdash 2 copy moveto
      2 copy vpt 270 360 arc closepath fill
              vpt 0 360 arc closepath } bind def
/C9 { BL [] 0 setdash 2 copy moveto
      2 copy  vpt 270 450 arc closepath fill
              vpt 0 360 arc closepath } bind def
/C10 { BL [] 0 setdash 2 copy 2 copy moveto vpt 270 360 arc closepath fill
       2 copy moveto
       2 copy vpt 90 180 arc closepath fill
               vpt 0 360 arc closepath } bind def
/C11 { BL [] 0 setdash 2 copy moveto
       2 copy  vpt 0 180 arc closepath fill
       2 copy moveto
       2 copy  vpt 270 360 arc closepath fill
               vpt 0 360 arc closepath } bind def
/C12 { BL [] 0 setdash 2 copy moveto
       2 copy  vpt 180 360 arc closepath fill
               vpt 0 360 arc closepath } bind def
/C13 { BL [] 0 setdash  2 copy moveto
       2 copy  vpt 0 90 arc closepath fill
       2 copy moveto
       2 copy  vpt 180 360 arc closepath fill
               vpt 0 360 arc closepath } bind def
/C14 { BL [] 0 setdash 2 copy moveto
       2 copy  vpt 90 360 arc closepath fill
               vpt 0 360 arc } bind def
/C15 { BL [] 0 setdash 2 copy vpt 0 360 arc closepath fill
               vpt 0 360 arc closepath } bind def
/Rec   { newpath 4 2 roll moveto 1 index 0 rlineto 0 exch rlineto
       neg 0 rlineto closepath } bind def
/Square { dup Rec } bind def
/Bsquare { vpt sub exch vpt sub exch vpt2 Square } bind def
/S0 { BL [] 0 setdash 2 copy moveto 0 vpt rlineto BL Bsquare } bind def
/S1 { BL [] 0 setdash 2 copy vpt Square fill Bsquare } bind def
/S2 { BL [] 0 setdash 2 copy exch vpt sub exch vpt Square fill Bsquare } bind def
/S3 { BL [] 0 setdash 2 copy exch vpt sub exch vpt2 vpt Rec fill Bsquare } bind def
/S4 { BL [] 0 setdash 2 copy exch vpt sub exch vpt sub vpt Square fill Bsquare } bind def
/S5 { BL [] 0 setdash 2 copy 2 copy vpt Square fill
       exch vpt sub exch vpt sub vpt Square fill Bsquare } bind def
/S6 { BL [] 0 setdash 2 copy exch vpt sub exch vpt sub vpt vpt2 Rec fill Bsquare } bind def
/S7 { BL [] 0 setdash 2 copy exch vpt sub exch vpt sub vpt vpt2 Rec fill
       2 copy vpt Square fill
       Bsquare } bind def
/S8 { BL [] 0 setdash 2 copy vpt sub vpt Square fill Bsquare } bind def
/S9 { BL [] 0 setdash 2 copy vpt sub vpt vpt2 Rec fill Bsquare } bind def
/S10 { BL [] 0 setdash 2 copy vpt sub vpt Square fill 2 copy exch vpt sub exch vpt Square fill
       Bsquare } bind def
/S11 { BL [] 0 setdash 2 copy vpt sub vpt Square fill 2 copy exch vpt sub exch vpt2 vpt Rec fill
       Bsquare } bind def
/S12 { BL [] 0 setdash 2 copy exch vpt sub exch vpt sub vpt2 vpt Rec fill Bsquare } bind def
/S13 { BL [] 0 setdash 2 copy exch vpt sub exch vpt sub vpt2 vpt Rec fill
       2 copy vpt Square fill Bsquare } bind def
/S14 { BL [] 0 setdash 2 copy exch vpt sub exch vpt sub vpt2 vpt Rec fill
       2 copy exch vpt sub exch vpt Square fill Bsquare } bind def
/S15 { BL [] 0 setdash 2 copy Bsquare fill Bsquare } bind def
/D0 { gsave translate 45 rotate 0 0 S0 stroke grestore } bind def
/D1 { gsave translate 45 rotate 0 0 S1 stroke grestore } bind def
/D2 { gsave translate 45 rotate 0 0 S2 stroke grestore } bind def
/D3 { gsave translate 45 rotate 0 0 S3 stroke grestore } bind def
/D4 { gsave translate 45 rotate 0 0 S4 stroke grestore } bind def
/D5 { gsave translate 45 rotate 0 0 S5 stroke grestore } bind def
/D6 { gsave translate 45 rotate 0 0 S6 stroke grestore } bind def
/D7 { gsave translate 45 rotate 0 0 S7 stroke grestore } bind def
/D8 { gsave translate 45 rotate 0 0 S8 stroke grestore } bind def
/D9 { gsave translate 45 rotate 0 0 S9 stroke grestore } bind def
/D10 { gsave translate 45 rotate 0 0 S10 stroke grestore } bind def
/D11 { gsave translate 45 rotate 0 0 S11 stroke grestore } bind def
/D12 { gsave translate 45 rotate 0 0 S12 stroke grestore } bind def
/D13 { gsave translate 45 rotate 0 0 S13 stroke grestore } bind def
/D14 { gsave translate 45 rotate 0 0 S14 stroke grestore } bind def
/D15 { gsave translate 45 rotate 0 0 S15 stroke grestore } bind def
/DiaE { stroke [] 0 setdash vpt add M
  hpt neg vpt neg V hpt vpt neg V
  hpt vpt V hpt neg vpt V closepath stroke } def
/BoxE { stroke [] 0 setdash exch hpt sub exch vpt add M
  0 vpt2 neg V hpt2 0 V 0 vpt2 V
  hpt2 neg 0 V closepath stroke } def
/TriUE { stroke [] 0 setdash vpt 1.12 mul add M
  hpt neg vpt -1.62 mul V
  hpt 2 mul 0 V
  hpt neg vpt 1.62 mul V closepath stroke } def
/TriDE { stroke [] 0 setdash vpt 1.12 mul sub M
  hpt neg vpt 1.62 mul V
  hpt 2 mul 0 V
  hpt neg vpt -1.62 mul V closepath stroke } def
/PentE { stroke [] 0 setdash gsave
  translate 0 hpt M 4 {72 rotate 0 hpt L} repeat
  closepath stroke grestore } def
/CircE { stroke [] 0 setdash 
  hpt 0 360 arc stroke } def
/Opaque { gsave closepath 1 setgray fill grestore 0 setgray closepath } def
/DiaW { stroke [] 0 setdash vpt add M
  hpt neg vpt neg V hpt vpt neg V
  hpt vpt V hpt neg vpt V Opaque stroke } def
/BoxW { stroke [] 0 setdash exch hpt sub exch vpt add M
  0 vpt2 neg V hpt2 0 V 0 vpt2 V
  hpt2 neg 0 V Opaque stroke } def
/TriUW { stroke [] 0 setdash vpt 1.12 mul add M
  hpt neg vpt -1.62 mul V
  hpt 2 mul 0 V
  hpt neg vpt 1.62 mul V Opaque stroke } def
/TriDW { stroke [] 0 setdash vpt 1.12 mul sub M
  hpt neg vpt 1.62 mul V
  hpt 2 mul 0 V
  hpt neg vpt -1.62 mul V Opaque stroke } def
/PentW { stroke [] 0 setdash gsave
  translate 0 hpt M 4 {72 rotate 0 hpt L} repeat
  Opaque stroke grestore } def
/CircW { stroke [] 0 setdash 
  hpt 0 360 arc Opaque stroke } def
/BoxFill { gsave Rec 1 setgray fill grestore } def
/BoxColFill {
  gsave Rec
  /Fillden exch def
  currentrgbcolor
  /ColB exch def /ColG exch def /ColR exch def
  /ColR ColR Fillden mul Fillden sub 1 add def
  /ColG ColG Fillden mul Fillden sub 1 add def
  /ColB ColB Fillden mul Fillden sub 1 add def
  ColR ColG ColB setrgbcolor
  fill grestore } def
%
% PostScript Level 1 Pattern Fill routine
% Usage: x y w h s a XX PatternFill
%	x,y = lower left corner of box to be filled
%	w,h = width and height of box
%	  a = angle in degrees between lines and x-axis
%	 XX = 0/1 for no/yes cross-hatch
%
/PatternFill { gsave /PFa [ 9 2 roll ] def
    PFa 0 get PFa 2 get 2 div add PFa 1 get PFa 3 get 2 div add translate
    PFa 2 get -2 div PFa 3 get -2 div PFa 2 get PFa 3 get Rec
    gsave 1 setgray fill grestore clip
    currentlinewidth 0.5 mul setlinewidth
    /PFs PFa 2 get dup mul PFa 3 get dup mul add sqrt def
    0 0 M PFa 5 get rotate PFs -2 div dup translate
	0 1 PFs PFa 4 get div 1 add floor cvi
	{ PFa 4 get mul 0 M 0 PFs V } for
    0 PFa 6 get ne {
	0 1 PFs PFa 4 get div 1 add floor cvi
	{ PFa 4 get mul 0 2 1 roll M PFs 0 V } for
    } if
    stroke grestore } def
/Symbol-Oblique /Symbol findfont [1 0 .167 1 0 0] makefont
dup length dict begin {1 index /FID eq {pop pop} {def} ifelse} forall
currentdict end definefont pop
end
gnudict begin
gsave
0 0 translate
0.100 0.100 scale
0 setgray
newpath
1.000 UL
LTb
400 200 M
63 0 V
977 0 R
-63 0 V
1.000 UL
LTb
400 460 M
63 0 V
977 0 R
-63 0 V
1.000 UL
LTb
400 720 M
63 0 V
977 0 R
-63 0 V
1.000 UL
LTb
400 980 M
63 0 V
977 0 R
-63 0 V
1.000 UL
LTb
697 200 M
0 63 V
0 717 R
0 -63 V
1.000 UL
LTb
1069 200 M
0 63 V
0 717 R
0 -63 V
1.000 UL
LTb
1440 200 M
0 63 V
0 717 R
0 -63 V
1.000 UL
LTb
1.000 UL
LTa
697 200 M
0 780 V
1.000 UL
LTb
400 200 M
1040 0 V
0 780 V
400 980 L
0 -780 V
0.600 UP
LTb
LTb
LTb
LTb
1.000 UL
LT0
1432 980 M
-7 -20 V
-25 -61 V
-24 -59 V
-25 -56 V
-24 -54 V
-25 -51 V
-25 -49 V
-24 -47 V
-25 -43 V
-25 -41 V
-24 -39 V
-25 -36 V
-25 -33 V
-24 -30 V
-25 -28 V
-24 -25 V
-24 -22 V
-24 -20 V
984 249 L
961 234 L
937 223 L
914 213 L
-23 -7 V
-22 -4 V
-23 -2 V
-22 1 V
-21 2 V
-21 5 V
-21 7 V
-20 9 V
-20 10 V
-19 13 V
-19 14 V
-18 15 V
-18 17 V
-16 18 V
-17 19 V
-15 20 V
-15 21 V
-14 21 V
-14 21 V
-12 22 V
-12 21 V
-11 21 V
-10 20 V
-9 20 V
-9 18 V
-7 18 V
-7 16 V
-5 14 V
-5 12 V
-4 11 V
-3 8 V
-2 5 V
0 3 V
0 -1 V
1 -3 V
2 -8 V
3 -12 V
5 -16 V
6 -22 V
8 -31 V
12 -46 V
0.600 UP
1.000 UL
LT6
846 200 CircleF
1.000 UL
LTb
400 200 M
1040 0 V
0 780 V
400 980 L
0 -780 V
0.600 UP
stroke
grestore
end
showpage
}}%
\put(771,304){\makebox(0,0)[l]{$\bar s_{\rm cav}$}}%
\put(868,824){\makebox(0,0)[l]{(d)}}%
\put(474,850){\makebox(0,0)[l]{$L_1$}}%
\put(1217,278){\makebox(0,0)[l]{$s_{\rm cav}$}}%
\put(1440,100){\makebox(0,0){ 0.1}}%
\put(1069,100){\makebox(0,0){ 0.05}}%
\put(697,100){\makebox(0,0){ 0}}%
\put(350,980){\makebox(0,0)[r]{ 0.03}}%
\put(350,720){\makebox(0,0)[r]{ 0.02}}%
\put(350,460){\makebox(0,0)[r]{ 0.01}}%
\put(350,200){\makebox(0,0)[r]{ 0}}%
\end{picture}%
\endgroup
 

%% file: errexpldpc.tex
% GNUPLOT: LaTeX picture with Postscript
\begingroup%
  \makeatletter%
  \newcommand{\GNUPLOTspecial}{%
    \@sanitize\catcode`\%=14\relax\special}%
  \setlength{\unitlength}{0.1bp}%
{\GNUPLOTspecial{!
%!PS-Adobe-2.0 EPSF-2.0
%%Title: errexpldpc.tex
%%Creator: gnuplot 3.7 patchlevel 1
%%CreationDate: Wed May 17 12:12:42 2006
%%DocumentFonts: 
%%BoundingBox: 0 0 360 259
%%Orientation: Landscape
%%EndComments
/gnudict 256 dict def
gnudict begin
/Color true def
/Solid false def
/gnulinewidth 5.000 def
/userlinewidth gnulinewidth def
/vshift -33 def
/dl {10 mul} def
/hpt_ 31.5 def
/vpt_ 31.5 def
/hpt hpt_ def
/vpt vpt_ def
/M {moveto} bind def
/L {lineto} bind def
/R {rmoveto} bind def
/V {rlineto} bind def
/vpt2 vpt 2 mul def
/hpt2 hpt 2 mul def
/Lshow { currentpoint stroke M
  0 vshift R show } def
/Rshow { currentpoint stroke M
  dup stringwidth pop neg vshift R show } def
/Cshow { currentpoint stroke M
  dup stringwidth pop -2 div vshift R show } def
/UP { dup vpt_ mul /vpt exch def hpt_ mul /hpt exch def
  /hpt2 hpt 2 mul def /vpt2 vpt 2 mul def } def
/DL { Color {setrgbcolor Solid {pop []} if 0 setdash }
 {pop pop pop Solid {pop []} if 0 setdash} ifelse } def
/BL { stroke userlinewidth 2 mul setlinewidth } def
/AL { stroke userlinewidth 2 div setlinewidth } def
/UL { dup gnulinewidth mul /userlinewidth exch def
      10 mul /udl exch def } def
/PL { stroke userlinewidth setlinewidth } def
/LTb { BL [] 0 0 0 DL } def
/LTa { AL [1 udl mul 2 udl mul] 0 setdash 0 0 0 setrgbcolor } def
/LT0 { PL [] 1 0 0 DL } def
/LT1 { PL [4 dl 2 dl] 0 1 0 DL } def
/LT2 { PL [2 dl 3 dl] 0 0 1 DL } def
/LT3 { PL [1 dl 1.5 dl] 1 0 1 DL } def
/LT4 { PL [5 dl 2 dl 1 dl 2 dl] 0 1 1 DL } def
/LT5 { PL [4 dl 3 dl 1 dl 3 dl] 1 1 0 DL } def
/LT6 { PL [2 dl 2 dl 2 dl 4 dl] 0 0 0 DL } def
/LT7 { PL [2 dl 2 dl 2 dl 2 dl 2 dl 4 dl] 1 0.3 0 DL } def
/LT8 { PL [2 dl 2 dl 2 dl 2 dl 2 dl 2 dl 2 dl 4 dl] 0.5 0.5 0.5 DL } def
/Pnt { stroke [] 0 setdash
   gsave 1 setlinecap M 0 0 V stroke grestore } def
/Dia { stroke [] 0 setdash 2 copy vpt add M
  hpt neg vpt neg V hpt vpt neg V
  hpt vpt V hpt neg vpt V closepath stroke
  Pnt } def
/Pls { stroke [] 0 setdash vpt sub M 0 vpt2 V
  currentpoint stroke M
  hpt neg vpt neg R hpt2 0 V stroke
  } def
/Box { stroke [] 0 setdash 2 copy exch hpt sub exch vpt add M
  0 vpt2 neg V hpt2 0 V 0 vpt2 V
  hpt2 neg 0 V closepath stroke
  Pnt } def
/Crs { stroke [] 0 setdash exch hpt sub exch vpt add M
  hpt2 vpt2 neg V currentpoint stroke M
  hpt2 neg 0 R hpt2 vpt2 V stroke } def
/TriU { stroke [] 0 setdash 2 copy vpt 1.12 mul add M
  hpt neg vpt -1.62 mul V
  hpt 2 mul 0 V
  hpt neg vpt 1.62 mul V closepath stroke
  Pnt  } def
/Star { 2 copy Pls Crs } def
/BoxF { stroke [] 0 setdash exch hpt sub exch vpt add M
  0 vpt2 neg V  hpt2 0 V  0 vpt2 V
  hpt2 neg 0 V  closepath fill } def
/TriUF { stroke [] 0 setdash vpt 1.12 mul add M
  hpt neg vpt -1.62 mul V
  hpt 2 mul 0 V
  hpt neg vpt 1.62 mul V closepath fill } def
/TriD { stroke [] 0 setdash 2 copy vpt 1.12 mul sub M
  hpt neg vpt 1.62 mul V
  hpt 2 mul 0 V
  hpt neg vpt -1.62 mul V closepath stroke
  Pnt  } def
/TriDF { stroke [] 0 setdash vpt 1.12 mul sub M
  hpt neg vpt 1.62 mul V
  hpt 2 mul 0 V
  hpt neg vpt -1.62 mul V closepath fill} def
/DiaF { stroke [] 0 setdash vpt add M
  hpt neg vpt neg V hpt vpt neg V
  hpt vpt V hpt neg vpt V closepath fill } def
/Pent { stroke [] 0 setdash 2 copy gsave
  translate 0 hpt M 4 {72 rotate 0 hpt L} repeat
  closepath stroke grestore Pnt } def
/PentF { stroke [] 0 setdash gsave
  translate 0 hpt M 4 {72 rotate 0 hpt L} repeat
  closepath fill grestore } def
/Circle { stroke [] 0 setdash 2 copy
  hpt 0 360 arc stroke Pnt } def
/CircleF { stroke [] 0 setdash hpt 0 360 arc fill } def
/C0 { BL [] 0 setdash 2 copy moveto vpt 90 450  arc } bind def
/C1 { BL [] 0 setdash 2 copy        moveto
       2 copy  vpt 0 90 arc closepath fill
               vpt 0 360 arc closepath } bind def
/C2 { BL [] 0 setdash 2 copy moveto
       2 copy  vpt 90 180 arc closepath fill
               vpt 0 360 arc closepath } bind def
/C3 { BL [] 0 setdash 2 copy moveto
       2 copy  vpt 0 180 arc closepath fill
               vpt 0 360 arc closepath } bind def
/C4 { BL [] 0 setdash 2 copy moveto
       2 copy  vpt 180 270 arc closepath fill
               vpt 0 360 arc closepath } bind def
/C5 { BL [] 0 setdash 2 copy moveto
       2 copy  vpt 0 90 arc
       2 copy moveto
       2 copy  vpt 180 270 arc closepath fill
               vpt 0 360 arc } bind def
/C6 { BL [] 0 setdash 2 copy moveto
      2 copy  vpt 90 270 arc closepath fill
              vpt 0 360 arc closepath } bind def
/C7 { BL [] 0 setdash 2 copy moveto
      2 copy  vpt 0 270 arc closepath fill
              vpt 0 360 arc closepath } bind def
/C8 { BL [] 0 setdash 2 copy moveto
      2 copy vpt 270 360 arc closepath fill
              vpt 0 360 arc closepath } bind def
/C9 { BL [] 0 setdash 2 copy moveto
      2 copy  vpt 270 450 arc closepath fill
              vpt 0 360 arc closepath } bind def
/C10 { BL [] 0 setdash 2 copy 2 copy moveto vpt 270 360 arc closepath fill
       2 copy moveto
       2 copy vpt 90 180 arc closepath fill
               vpt 0 360 arc closepath } bind def
/C11 { BL [] 0 setdash 2 copy moveto
       2 copy  vpt 0 180 arc closepath fill
       2 copy moveto
       2 copy  vpt 270 360 arc closepath fill
               vpt 0 360 arc closepath } bind def
/C12 { BL [] 0 setdash 2 copy moveto
       2 copy  vpt 180 360 arc closepath fill
               vpt 0 360 arc closepath } bind def
/C13 { BL [] 0 setdash  2 copy moveto
       2 copy  vpt 0 90 arc closepath fill
       2 copy moveto
       2 copy  vpt 180 360 arc closepath fill
               vpt 0 360 arc closepath } bind def
/C14 { BL [] 0 setdash 2 copy moveto
       2 copy  vpt 90 360 arc closepath fill
               vpt 0 360 arc } bind def
/C15 { BL [] 0 setdash 2 copy vpt 0 360 arc closepath fill
               vpt 0 360 arc closepath } bind def
/Rec   { newpath 4 2 roll moveto 1 index 0 rlineto 0 exch rlineto
       neg 0 rlineto closepath } bind def
/Square { dup Rec } bind def
/Bsquare { vpt sub exch vpt sub exch vpt2 Square } bind def
/S0 { BL [] 0 setdash 2 copy moveto 0 vpt rlineto BL Bsquare } bind def
/S1 { BL [] 0 setdash 2 copy vpt Square fill Bsquare } bind def
/S2 { BL [] 0 setdash 2 copy exch vpt sub exch vpt Square fill Bsquare } bind def
/S3 { BL [] 0 setdash 2 copy exch vpt sub exch vpt2 vpt Rec fill Bsquare } bind def
/S4 { BL [] 0 setdash 2 copy exch vpt sub exch vpt sub vpt Square fill Bsquare } bind def
/S5 { BL [] 0 setdash 2 copy 2 copy vpt Square fill
       exch vpt sub exch vpt sub vpt Square fill Bsquare } bind def
/S6 { BL [] 0 setdash 2 copy exch vpt sub exch vpt sub vpt vpt2 Rec fill Bsquare } bind def
/S7 { BL [] 0 setdash 2 copy exch vpt sub exch vpt sub vpt vpt2 Rec fill
       2 copy vpt Square fill
       Bsquare } bind def
/S8 { BL [] 0 setdash 2 copy vpt sub vpt Square fill Bsquare } bind def
/S9 { BL [] 0 setdash 2 copy vpt sub vpt vpt2 Rec fill Bsquare } bind def
/S10 { BL [] 0 setdash 2 copy vpt sub vpt Square fill 2 copy exch vpt sub exch vpt Square fill
       Bsquare } bind def
/S11 { BL [] 0 setdash 2 copy vpt sub vpt Square fill 2 copy exch vpt sub exch vpt2 vpt Rec fill
       Bsquare } bind def
/S12 { BL [] 0 setdash 2 copy exch vpt sub exch vpt sub vpt2 vpt Rec fill Bsquare } bind def
/S13 { BL [] 0 setdash 2 copy exch vpt sub exch vpt sub vpt2 vpt Rec fill
       2 copy vpt Square fill Bsquare } bind def
/S14 { BL [] 0 setdash 2 copy exch vpt sub exch vpt sub vpt2 vpt Rec fill
       2 copy exch vpt sub exch vpt Square fill Bsquare } bind def
/S15 { BL [] 0 setdash 2 copy Bsquare fill Bsquare } bind def
/D0 { gsave translate 45 rotate 0 0 S0 stroke grestore } bind def
/D1 { gsave translate 45 rotate 0 0 S1 stroke grestore } bind def
/D2 { gsave translate 45 rotate 0 0 S2 stroke grestore } bind def
/D3 { gsave translate 45 rotate 0 0 S3 stroke grestore } bind def
/D4 { gsave translate 45 rotate 0 0 S4 stroke grestore } bind def
/D5 { gsave translate 45 rotate 0 0 S5 stroke grestore } bind def
/D6 { gsave translate 45 rotate 0 0 S6 stroke grestore } bind def
/D7 { gsave translate 45 rotate 0 0 S7 stroke grestore } bind def
/D8 { gsave translate 45 rotate 0 0 S8 stroke grestore } bind def
/D9 { gsave translate 45 rotate 0 0 S9 stroke grestore } bind def
/D10 { gsave translate 45 rotate 0 0 S10 stroke grestore } bind def
/D11 { gsave translate 45 rotate 0 0 S11 stroke grestore } bind def
/D12 { gsave translate 45 rotate 0 0 S12 stroke grestore } bind def
/D13 { gsave translate 45 rotate 0 0 S13 stroke grestore } bind def
/D14 { gsave translate 45 rotate 0 0 S14 stroke grestore } bind def
/D15 { gsave translate 45 rotate 0 0 S15 stroke grestore } bind def
/DiaE { stroke [] 0 setdash vpt add M
  hpt neg vpt neg V hpt vpt neg V
  hpt vpt V hpt neg vpt V closepath stroke } def
/BoxE { stroke [] 0 setdash exch hpt sub exch vpt add M
  0 vpt2 neg V hpt2 0 V 0 vpt2 V
  hpt2 neg 0 V closepath stroke } def
/TriUE { stroke [] 0 setdash vpt 1.12 mul add M
  hpt neg vpt -1.62 mul V
  hpt 2 mul 0 V
  hpt neg vpt 1.62 mul V closepath stroke } def
/TriDE { stroke [] 0 setdash vpt 1.12 mul sub M
  hpt neg vpt 1.62 mul V
  hpt 2 mul 0 V
  hpt neg vpt -1.62 mul V closepath stroke } def
/PentE { stroke [] 0 setdash gsave
  translate 0 hpt M 4 {72 rotate 0 hpt L} repeat
  closepath stroke grestore } def
/CircE { stroke [] 0 setdash 
  hpt 0 360 arc stroke } def
/Opaque { gsave closepath 1 setgray fill grestore 0 setgray closepath } def
/DiaW { stroke [] 0 setdash vpt add M
  hpt neg vpt neg V hpt vpt neg V
  hpt vpt V hpt neg vpt V Opaque stroke } def
/BoxW { stroke [] 0 setdash exch hpt sub exch vpt add M
  0 vpt2 neg V hpt2 0 V 0 vpt2 V
  hpt2 neg 0 V Opaque stroke } def
/TriUW { stroke [] 0 setdash vpt 1.12 mul add M
  hpt neg vpt -1.62 mul V
  hpt 2 mul 0 V
  hpt neg vpt 1.62 mul V Opaque stroke } def
/TriDW { stroke [] 0 setdash vpt 1.12 mul sub M
  hpt neg vpt 1.62 mul V
  hpt 2 mul 0 V
  hpt neg vpt -1.62 mul V Opaque stroke } def
/PentW { stroke [] 0 setdash gsave
  translate 0 hpt M 4 {72 rotate 0 hpt L} repeat
  Opaque stroke grestore } def
/CircW { stroke [] 0 setdash 
  hpt 0 360 arc Opaque stroke } def
/BoxFill { gsave Rec 1 setgray fill grestore } def
end
}}%
\begin{picture}(3600,2592)(0,0)%
{\GNUPLOTspecial{"
gnudict begin
gsave
0 0 translate
0.100 0.100 scale
0 setgray
newpath
1.000 UL
LTb
450 300 M
63 0 V
2937 0 R
-63 0 V
450 553 M
63 0 V
2937 0 R
-63 0 V
450 806 M
63 0 V
2937 0 R
-63 0 V
450 1060 M
63 0 V
2937 0 R
-63 0 V
450 1313 M
63 0 V
2937 0 R
-63 0 V
450 1566 M
63 0 V
2937 0 R
-63 0 V
450 1819 M
63 0 V
2937 0 R
-63 0 V
450 2073 M
63 0 V
2937 0 R
-63 0 V
450 2326 M
63 0 V
2937 0 R
-63 0 V
450 300 M
0 63 V
0 2129 R
0 -63 V
783 300 M
0 63 V
0 2129 R
0 -63 V
1117 300 M
0 63 V
0 2129 R
0 -63 V
1450 300 M
0 63 V
0 2129 R
0 -63 V
1783 300 M
0 63 V
0 2129 R
0 -63 V
2117 300 M
0 63 V
0 2129 R
0 -63 V
2450 300 M
0 63 V
0 2129 R
0 -63 V
2783 300 M
0 63 V
0 2129 R
0 -63 V
3117 300 M
0 63 V
0 2129 R
0 -63 V
3450 300 M
0 63 V
0 2129 R
0 -63 V
1.000 UL
LTb
450 300 M
3000 0 V
0 2192 V
-3000 0 V
450 300 L
1.000 UL
LT0
3106 2379 M
244 0 V
3236 300 M
-17 0 V
-16 1 V
-17 0 V
-17 1 V
-16 0 V
-17 1 V
-17 1 V
-16 2 V
-17 1 V
-17 1 V
-16 2 V
-17 2 V
-17 2 V
-16 2 V
-17 2 V
-17 2 V
-16 2 V
-17 3 V
-17 3 V
-16 3 V
-17 3 V
-17 3 V
-16 3 V
-17 3 V
-17 4 V
-16 4 V
-17 3 V
-17 4 V
-16 4 V
-17 5 V
-17 4 V
-16 5 V
-17 4 V
-17 5 V
-16 5 V
-17 5 V
-17 5 V
-16 6 V
-17 5 V
-17 6 V
-16 6 V
-17 5 V
-17 7 V
-16 6 V
-17 6 V
-17 7 V
-16 6 V
-17 7 V
-17 7 V
-16 7 V
-17 7 V
-17 7 V
-16 8 V
-17 8 V
-17 7 V
-16 8 V
-17 8 V
-17 9 V
-16 8 V
-17 8 V
-17 9 V
-16 9 V
-17 9 V
-17 9 V
-16 9 V
-17 9 V
-17 10 V
-16 9 V
-17 10 V
-17 10 V
-16 10 V
-17 10 V
-17 11 V
-16 10 V
-17 11 V
-17 11 V
-16 11 V
-17 11 V
-17 11 V
-16 11 V
-17 12 V
-17 12 V
-16 12 V
-17 12 V
-17 12 V
-16 12 V
-17 12 V
-17 13 V
-16 13 V
-17 13 V
-17 13 V
-16 13 V
-17 13 V
-17 14 V
-16 13 V
-17 14 V
-17 14 V
-16 14 V
-17 14 V
-17 15 V
-16 14 V
-17 15 V
-17 15 V
-16 15 V
-17 15 V
-17 15 V
-16 16 V
-17 16 V
-17 15 V
-16 16 V
-17 16 V
-17 17 V
-16 16 V
-17 17 V
-17 17 V
-16 16 V
-17 18 V
-17 17 V
-16 17 V
-17 18 V
-17 17 V
-16 18 V
-17 18 V
-17 18 V
-16 19 V
-17 18 V
-17 19 V
-16 19 V
-17 19 V
-17 19 V
-16 19 V
-17 20 V
-17 20 V
-16 19 V
-17 20 V
-17 21 V
-16 20 V
-16 19 V
-2 3 V
-2 2 V
-1 2 V
-2 2 V
-2 2 V
-1 2 V
-2 2 V
-2 2 V
-1 2 V
-2 2 V
-2 2 V
-1 2 V
-2 2 V
-2 2 V
-1 2 V
-2 2 V
-2 2 V
-1 2 V
-2 2 V
-2 2 V
-1 2 V
-2 2 V
-2 2 V
-1 2 V
-2 2 V
-2 2 V
-1 2 V
-2 2 V
-2 2 V
-1 2 V
-2 2 V
-2 2 V
-1 2 V
-2 2 V
-2 2 V
-1 2 V
-2 2 V
-2 2 V
-1 2 V
-2 2 V
-2 2 V
-1 2 V
-2 2 V
-2 2 V
-1 2 V
-2 2 V
-2 2 V
-1 2 V
-2 2 V
-2 2 V
-1 2 V
-2 2 V
-2 2 V
-1 2 V
-2 2 V
-2 2 V
-1 2 V
-2 2 V
-2 2 V
-1 2 V
-2 2 V
-2 2 V
-1 1 V
-2 2 V
-2 2 V
-1 2 V
-2 2 V
-2 2 V
-1 2 V
-2 2 V
-2 2 V
-1 2 V
-2 2 V
-2 2 V
-1 1 V
-2 2 V
-2 2 V
-1 2 V
-2 2 V
-2 2 V
-1 2 V
-2 2 V
-2 1 V
-1 2 V
-2 2 V
-2 2 V
-1 2 V
-2 2 V
-2 2 V
-1 1 V
-2 2 V
-2 2 V
-1 2 V
-2 2 V
-2 2 V
-1 1 V
-2 2 V
-2 2 V
-1 2 V
-2 2 V
-2 1 V
-1 2 V
-2 2 V
-2 2 V
-1 2 V
-2 1 V
-2 2 V
-1 2 V
-2 1 V
-2 2 V
-1 2 V
-2 2 V
-2 1 V
0.700 UP
1.000 UL
LTa
937 1682 CircleF
0.700 UP
1.000 UL
LTa
748 1900 CircleF
0.700 UP
1.000 UL
LT6
3252 300 CircleF
1.000 UL
LT1
3106 2279 M
244 0 V
450 1246 M
17 -2 V
16 -3 V
17 -2 V
17 -3 V
16 -3 V
17 -2 V
17 -3 V
16 -2 V
17 -3 V
17 -2 V
16 -3 V
17 -2 V
17 -3 V
16 -2 V
17 -3 V
17 -2 V
16 -3 V
17 -2 V
17 -3 V
16 -2 V
17 -2 V
17 -3 V
16 -2 V
17 -3 V
17 -2 V
16 -3 V
17 -2 V
17 -2 V
16 -3 V
17 -2 V
17 -2 V
16 -3 V
17 -2 V
17 -3 V
16 -2 V
17 -2 V
17 -3 V
16 -2 V
17 -2 V
17 -2 V
16 -3 V
17 -2 V
17 -2 V
16 -3 V
17 -2 V
17 -2 V
16 -2 V
17 -3 V
17 -2 V
16 -2 V
17 -2 V
17 -3 V
16 -2 V
17 -2 V
17 -2 V
16 -3 V
17 -2 V
17 -2 V
16 -2 V
17 -2 V
3 -1 V
17 -16 V
17 -15 V
16 -16 V
17 -15 V
17 -16 V
16 -15 V
17 -15 V
17 -14 V
16 -15 V
17 -14 V
17 -15 V
16 -14 V
17 -13 V
17 -14 V
16 -14 V
17 -13 V
17 -13 V
16 -13 V
17 -13 V
17 -13 V
16 -13 V
17 -12 V
17 -12 V
16 -12 V
17 -12 V
17 -12 V
16 -12 V
17 -11 V
17 -11 V
16 -12 V
17 -11 V
17 -10 V
16 -11 V
17 -10 V
17 -11 V
16 -10 V
17 -10 V
17 -10 V
16 -10 V
17 -9 V
17 -10 V
16 -9 V
17 -9 V
17 -9 V
16 -9 V
17 -8 V
17 -9 V
16 -8 V
17 -8 V
17 -8 V
16 -8 V
17 -8 V
17 -7 V
16 -8 V
17 -7 V
17 -7 V
16 -7 V
17 -7 V
17 -6 V
16 -7 V
17 -6 V
17 -6 V
16 -6 V
17 -6 V
17 -6 V
16 -5 V
17 -6 V
17 -5 V
16 -5 V
17 -5 V
17 -5 V
16 -5 V
17 -4 V
17 -5 V
16 -4 V
17 -4 V
17 -4 V
16 -4 V
17 -3 V
17 -4 V
16 -3 V
17 -3 V
17 -3 V
16 -3 V
17 -3 V
17 -2 V
16 -3 V
17 -2 V
17 -2 V
16 -2 V
17 -2 V
17 -2 V
16 -1 V
17 -2 V
17 -1 V
16 -1 V
17 -1 V
17 -1 V
16 -1 V
17 0 V
17 -1 V
16 0 V
17 0 V
1.000 UL
LT2
3106 2179 M
244 0 V
648 2492 M
2 -4 V
17 -27 V
16 -28 V
17 -27 V
17 -27 V
16 -27 V
17 -26 V
17 -26 V
16 -27 V
17 -26 V
17 -25 V
16 -26 V
17 -25 V
17 -25 V
16 -25 V
17 -25 V
17 -25 V
16 -24 V
17 -24 V
17 -24 V
16 -24 V
17 -24 V
17 -23 V
16 -23 V
17 -23 V
17 -23 V
16 -23 V
17 -22 V
17 -23 V
16 -22 V
17 -22 V
17 -21 V
16 -22 V
17 -21 V
17 -22 V
16 -21 V
17 -20 V
17 -21 V
16 -20 V
17 -21 V
17 -20 V
16 -20 V
17 -20 V
17 -19 V
16 -19 V
17 -20 V
17 -19 V
16 -19 V
17 -18 V
17 -19 V
16 -18 V
17 -18 V
17 -18 V
16 -18 V
17 -18 V
17 -17 V
16 -18 V
17 -17 V
17 -17 V
16 -17 V
17 -16 V
17 -17 V
16 -16 V
17 -16 V
17 -16 V
16 -16 V
17 -16 V
17 -15 V
16 -15 V
17 -15 V
17 -15 V
16 -15 V
17 -15 V
17 -14 V
16 -15 V
17 -14 V
17 -14 V
16 -14 V
17 -13 V
17 -14 V
16 -13 V
17 -13 V
17 -13 V
16 -13 V
17 -13 V
17 -13 V
16 -12 V
17 -12 V
17 -12 V
16 -12 V
17 -12 V
17 -12 V
16 -11 V
17 -11 V
17 -11 V
16 -11 V
17 -11 V
17 -11 V
16 -10 V
17 -11 V
17 -10 V
16 -10 V
17 -10 V
17 -10 V
16 -9 V
17 -10 V
17 -9 V
16 -9 V
17 -9 V
17 -9 V
16 -9 V
17 -8 V
17 -8 V
16 -9 V
17 -8 V
17 -8 V
16 -7 V
17 -8 V
17 -7 V
16 -8 V
17 -7 V
17 -7 V
16 -7 V
17 -6 V
17 -7 V
16 -6 V
17 -7 V
17 -6 V
16 -6 V
17 -6 V
17 -5 V
16 -6 V
17 -5 V
17 -5 V
16 -5 V
17 -5 V
17 -5 V
16 -5 V
17 -4 V
17 -4 V
16 -5 V
17 -4 V
17 -4 V
16 -3 V
17 -4 V
17 -3 V
16 -4 V
17 -3 V
17 -3 V
16 -3 V
17 -2 V
17 -3 V
16 -2 V
17 -3 V
17 -2 V
16 -2 V
17 -1 V
17 -2 V
16 -2 V
17 -1 V
17 -1 V
16 -1 V
17 -1 V
17 -1 V
16 -1 V
17 0 V
17 -1 V
16 0 V
17 0 V
0.700 UP
1.000 UL
LT6
672 2452 CircleF
3450 300 CircleF
stroke
grestore
end
showpage
}}%
\put(3056,2179){\makebox(0,0)[r]{Random Linear Model}}%
\put(3056,2279){\makebox(0,0)[r]{Union Bound}}%
\put(3056,2379){\makebox(0,0)[r]{$E_{\rm av}$}}%
\put(617,2326){\makebox(0,0)[l]{$p_e({\rm RLC})$}}%
\put(783,1944){\makebox(0,0)[l]{$p_{\rm rs}$}}%
\put(983,1725){\makebox(0,0)[l]{$p_e$}}%
\put(3200,391){\makebox(0,0)[l]{$p_c$}}%
\put(1950,50){\makebox(0,0){$p$}}%
\put(100,1396){%
\special{ps: gsave currentpoint currentpoint translate
270 rotate neg exch neg exch translate}%
\makebox(0,0)[b]{\shortstack{$E$}}%
\special{ps: currentpoint grestore moveto}%
}%
\put(3450,200){\makebox(0,0){0.5}}%
\put(3117,200){\makebox(0,0){0.48}}%
\put(2783,200){\makebox(0,0){0.46}}%
\put(2450,200){\makebox(0,0){0.44}}%
\put(2117,200){\makebox(0,0){0.42}}%
\put(1783,200){\makebox(0,0){0.4}}%
\put(1450,200){\makebox(0,0){0.38}}%
\put(1117,200){\makebox(0,0){0.36}}%
\put(783,200){\makebox(0,0){0.34}}%
\put(450,200){\makebox(0,0){0.32}}%
\put(400,2326){\makebox(0,0)[r]{0.08}}%
\put(400,2073){\makebox(0,0)[r]{0.07}}%
\put(400,1819){\makebox(0,0)[r]{0.06}}%
\put(400,1566){\makebox(0,0)[r]{0.05}}%
\put(400,1313){\makebox(0,0)[r]{0.04}}%
\put(400,1060){\makebox(0,0)[r]{0.03}}%
\put(400,806){\makebox(0,0)[r]{0.02}}%
\put(400,553){\makebox(0,0)[r]{0.01}}%
\put(400,300){\makebox(0,0)[r]{0}}%
\end{picture}%
\endgroup
 

%% file: rlmbec.tex
% GNUPLOT: LaTeX picture with Postscript
\begingroup%
  \makeatletter%
  \newcommand{\GNUPLOTspecial}{%
    \@sanitize\catcode`\%=14\relax\special}%
  \setlength{\unitlength}{0.1bp}%
{\GNUPLOTspecial{!
%!PS-Adobe-2.0 EPSF-2.0
%%Title: rlmbec.tex
%%Creator: gnuplot 3.7 patchlevel 1
%%CreationDate: Wed May  3 10:54:02 2006
%%DocumentFonts: 
%%BoundingBox: 0 0 360 216
%%Orientation: Landscape
%%EndComments
/gnudict 256 dict def
gnudict begin
/Color false def
/Solid false def
/gnulinewidth 5.000 def
/userlinewidth gnulinewidth def
/vshift -33 def
/dl {10 mul} def
/hpt_ 31.5 def
/vpt_ 31.5 def
/hpt hpt_ def
/vpt vpt_ def
/M {moveto} bind def
/L {lineto} bind def
/R {rmoveto} bind def
/V {rlineto} bind def
/vpt2 vpt 2 mul def
/hpt2 hpt 2 mul def
/Lshow { currentpoint stroke M
  0 vshift R show } def
/Rshow { currentpoint stroke M
  dup stringwidth pop neg vshift R show } def
/Cshow { currentpoint stroke M
  dup stringwidth pop -2 div vshift R show } def
/UP { dup vpt_ mul /vpt exch def hpt_ mul /hpt exch def
  /hpt2 hpt 2 mul def /vpt2 vpt 2 mul def } def
/DL { Color {setrgbcolor Solid {pop []} if 0 setdash }
 {pop pop pop Solid {pop []} if 0 setdash} ifelse } def
/BL { stroke userlinewidth 2 mul setlinewidth } def
/AL { stroke userlinewidth 2 div setlinewidth } def
/UL { dup gnulinewidth mul /userlinewidth exch def
      10 mul /udl exch def } def
/PL { stroke userlinewidth setlinewidth } def
/LTb { BL [] 0 0 0 DL } def
/LTa { AL [1 udl mul 2 udl mul] 0 setdash 0 0 0 setrgbcolor } def
/LT0 { PL [] 1 0 0 DL } def
/LT1 { PL [4 dl 2 dl] 0 1 0 DL } def
/LT2 { PL [2 dl 3 dl] 0 0 1 DL } def
/LT3 { PL [1 dl 1.5 dl] 1 0 1 DL } def
/LT4 { PL [5 dl 2 dl 1 dl 2 dl] 0 1 1 DL } def
/LT5 { PL [4 dl 3 dl 1 dl 3 dl] 1 1 0 DL } def
/LT6 { PL [2 dl 2 dl 2 dl 4 dl] 0 0 0 DL } def
/LT7 { PL [2 dl 2 dl 2 dl 2 dl 2 dl 4 dl] 1 0.3 0 DL } def
/LT8 { PL [2 dl 2 dl 2 dl 2 dl 2 dl 2 dl 2 dl 4 dl] 0.5 0.5 0.5 DL } def
/Pnt { stroke [] 0 setdash
   gsave 1 setlinecap M 0 0 V stroke grestore } def
/Dia { stroke [] 0 setdash 2 copy vpt add M
  hpt neg vpt neg V hpt vpt neg V
  hpt vpt V hpt neg vpt V closepath stroke
  Pnt } def
/Pls { stroke [] 0 setdash vpt sub M 0 vpt2 V
  currentpoint stroke M
  hpt neg vpt neg R hpt2 0 V stroke
  } def
/Box { stroke [] 0 setdash 2 copy exch hpt sub exch vpt add M
  0 vpt2 neg V hpt2 0 V 0 vpt2 V
  hpt2 neg 0 V closepath stroke
  Pnt } def
/Crs { stroke [] 0 setdash exch hpt sub exch vpt add M
  hpt2 vpt2 neg V currentpoint stroke M
  hpt2 neg 0 R hpt2 vpt2 V stroke } def
/TriU { stroke [] 0 setdash 2 copy vpt 1.12 mul add M
  hpt neg vpt -1.62 mul V
  hpt 2 mul 0 V
  hpt neg vpt 1.62 mul V closepath stroke
  Pnt  } def
/Star { 2 copy Pls Crs } def
/BoxF { stroke [] 0 setdash exch hpt sub exch vpt add M
  0 vpt2 neg V  hpt2 0 V  0 vpt2 V
  hpt2 neg 0 V  closepath fill } def
/TriUF { stroke [] 0 setdash vpt 1.12 mul add M
  hpt neg vpt -1.62 mul V
  hpt 2 mul 0 V
  hpt neg vpt 1.62 mul V closepath fill } def
/TriD { stroke [] 0 setdash 2 copy vpt 1.12 mul sub M
  hpt neg vpt 1.62 mul V
  hpt 2 mul 0 V
  hpt neg vpt -1.62 mul V closepath stroke
  Pnt  } def
/TriDF { stroke [] 0 setdash vpt 1.12 mul sub M
  hpt neg vpt 1.62 mul V
  hpt 2 mul 0 V
  hpt neg vpt -1.62 mul V closepath fill} def
/DiaF { stroke [] 0 setdash vpt add M
  hpt neg vpt neg V hpt vpt neg V
  hpt vpt V hpt neg vpt V closepath fill } def
/Pent { stroke [] 0 setdash 2 copy gsave
  translate 0 hpt M 4 {72 rotate 0 hpt L} repeat
  closepath stroke grestore Pnt } def
/PentF { stroke [] 0 setdash gsave
  translate 0 hpt M 4 {72 rotate 0 hpt L} repeat
  closepath fill grestore } def
/Circle { stroke [] 0 setdash 2 copy
  hpt 0 360 arc stroke Pnt } def
/CircleF { stroke [] 0 setdash hpt 0 360 arc fill } def
/C0 { BL [] 0 setdash 2 copy moveto vpt 90 450  arc } bind def
/C1 { BL [] 0 setdash 2 copy        moveto
       2 copy  vpt 0 90 arc closepath fill
               vpt 0 360 arc closepath } bind def
/C2 { BL [] 0 setdash 2 copy moveto
       2 copy  vpt 90 180 arc closepath fill
               vpt 0 360 arc closepath } bind def
/C3 { BL [] 0 setdash 2 copy moveto
       2 copy  vpt 0 180 arc closepath fill
               vpt 0 360 arc closepath } bind def
/C4 { BL [] 0 setdash 2 copy moveto
       2 copy  vpt 180 270 arc closepath fill
               vpt 0 360 arc closepath } bind def
/C5 { BL [] 0 setdash 2 copy moveto
       2 copy  vpt 0 90 arc
       2 copy moveto
       2 copy  vpt 180 270 arc closepath fill
               vpt 0 360 arc } bind def
/C6 { BL [] 0 setdash 2 copy moveto
      2 copy  vpt 90 270 arc closepath fill
              vpt 0 360 arc closepath } bind def
/C7 { BL [] 0 setdash 2 copy moveto
      2 copy  vpt 0 270 arc closepath fill
              vpt 0 360 arc closepath } bind def
/C8 { BL [] 0 setdash 2 copy moveto
      2 copy vpt 270 360 arc closepath fill
              vpt 0 360 arc closepath } bind def
/C9 { BL [] 0 setdash 2 copy moveto
      2 copy  vpt 270 450 arc closepath fill
              vpt 0 360 arc closepath } bind def
/C10 { BL [] 0 setdash 2 copy 2 copy moveto vpt 270 360 arc closepath fill
       2 copy moveto
       2 copy vpt 90 180 arc closepath fill
               vpt 0 360 arc closepath } bind def
/C11 { BL [] 0 setdash 2 copy moveto
       2 copy  vpt 0 180 arc closepath fill
       2 copy moveto
       2 copy  vpt 270 360 arc closepath fill
               vpt 0 360 arc closepath } bind def
/C12 { BL [] 0 setdash 2 copy moveto
       2 copy  vpt 180 360 arc closepath fill
               vpt 0 360 arc closepath } bind def
/C13 { BL [] 0 setdash  2 copy moveto
       2 copy  vpt 0 90 arc closepath fill
       2 copy moveto
       2 copy  vpt 180 360 arc closepath fill
               vpt 0 360 arc closepath } bind def
/C14 { BL [] 0 setdash 2 copy moveto
       2 copy  vpt 90 360 arc closepath fill
               vpt 0 360 arc } bind def
/C15 { BL [] 0 setdash 2 copy vpt 0 360 arc closepath fill
               vpt 0 360 arc closepath } bind def
/Rec   { newpath 4 2 roll moveto 1 index 0 rlineto 0 exch rlineto
       neg 0 rlineto closepath } bind def
/Square { dup Rec } bind def
/Bsquare { vpt sub exch vpt sub exch vpt2 Square } bind def
/S0 { BL [] 0 setdash 2 copy moveto 0 vpt rlineto BL Bsquare } bind def
/S1 { BL [] 0 setdash 2 copy vpt Square fill Bsquare } bind def
/S2 { BL [] 0 setdash 2 copy exch vpt sub exch vpt Square fill Bsquare } bind def
/S3 { BL [] 0 setdash 2 copy exch vpt sub exch vpt2 vpt Rec fill Bsquare } bind def
/S4 { BL [] 0 setdash 2 copy exch vpt sub exch vpt sub vpt Square fill Bsquare } bind def
/S5 { BL [] 0 setdash 2 copy 2 copy vpt Square fill
       exch vpt sub exch vpt sub vpt Square fill Bsquare } bind def
/S6 { BL [] 0 setdash 2 copy exch vpt sub exch vpt sub vpt vpt2 Rec fill Bsquare } bind def
/S7 { BL [] 0 setdash 2 copy exch vpt sub exch vpt sub vpt vpt2 Rec fill
       2 copy vpt Square fill
       Bsquare } bind def
/S8 { BL [] 0 setdash 2 copy vpt sub vpt Square fill Bsquare } bind def
/S9 { BL [] 0 setdash 2 copy vpt sub vpt vpt2 Rec fill Bsquare } bind def
/S10 { BL [] 0 setdash 2 copy vpt sub vpt Square fill 2 copy exch vpt sub exch vpt Square fill
       Bsquare } bind def
/S11 { BL [] 0 setdash 2 copy vpt sub vpt Square fill 2 copy exch vpt sub exch vpt2 vpt Rec fill
       Bsquare } bind def
/S12 { BL [] 0 setdash 2 copy exch vpt sub exch vpt sub vpt2 vpt Rec fill Bsquare } bind def
/S13 { BL [] 0 setdash 2 copy exch vpt sub exch vpt sub vpt2 vpt Rec fill
       2 copy vpt Square fill Bsquare } bind def
/S14 { BL [] 0 setdash 2 copy exch vpt sub exch vpt sub vpt2 vpt Rec fill
       2 copy exch vpt sub exch vpt Square fill Bsquare } bind def
/S15 { BL [] 0 setdash 2 copy Bsquare fill Bsquare } bind def
/D0 { gsave translate 45 rotate 0 0 S0 stroke grestore } bind def
/D1 { gsave translate 45 rotate 0 0 S1 stroke grestore } bind def
/D2 { gsave translate 45 rotate 0 0 S2 stroke grestore } bind def
/D3 { gsave translate 45 rotate 0 0 S3 stroke grestore } bind def
/D4 { gsave translate 45 rotate 0 0 S4 stroke grestore } bind def
/D5 { gsave translate 45 rotate 0 0 S5 stroke grestore } bind def
/D6 { gsave translate 45 rotate 0 0 S6 stroke grestore } bind def
/D7 { gsave translate 45 rotate 0 0 S7 stroke grestore } bind def
/D8 { gsave translate 45 rotate 0 0 S8 stroke grestore } bind def
/D9 { gsave translate 45 rotate 0 0 S9 stroke grestore } bind def
/D10 { gsave translate 45 rotate 0 0 S10 stroke grestore } bind def
/D11 { gsave translate 45 rotate 0 0 S11 stroke grestore } bind def
/D12 { gsave translate 45 rotate 0 0 S12 stroke grestore } bind def
/D13 { gsave translate 45 rotate 0 0 S13 stroke grestore } bind def
/D14 { gsave translate 45 rotate 0 0 S14 stroke grestore } bind def
/D15 { gsave translate 45 rotate 0 0 S15 stroke grestore } bind def
/DiaE { stroke [] 0 setdash vpt add M
  hpt neg vpt neg V hpt vpt neg V
  hpt vpt V hpt neg vpt V closepath stroke } def
/BoxE { stroke [] 0 setdash exch hpt sub exch vpt add M
  0 vpt2 neg V hpt2 0 V 0 vpt2 V
  hpt2 neg 0 V closepath stroke } def
/TriUE { stroke [] 0 setdash vpt 1.12 mul add M
  hpt neg vpt -1.62 mul V
  hpt 2 mul 0 V
  hpt neg vpt 1.62 mul V closepath stroke } def
/TriDE { stroke [] 0 setdash vpt 1.12 mul sub M
  hpt neg vpt 1.62 mul V
  hpt 2 mul 0 V
  hpt neg vpt -1.62 mul V closepath stroke } def
/PentE { stroke [] 0 setdash gsave
  translate 0 hpt M 4 {72 rotate 0 hpt L} repeat
  closepath stroke grestore } def
/CircE { stroke [] 0 setdash 
  hpt 0 360 arc stroke } def
/Opaque { gsave closepath 1 setgray fill grestore 0 setgray closepath } def
/DiaW { stroke [] 0 setdash vpt add M
  hpt neg vpt neg V hpt vpt neg V
  hpt vpt V hpt neg vpt V Opaque stroke } def
/BoxW { stroke [] 0 setdash exch hpt sub exch vpt add M
  0 vpt2 neg V hpt2 0 V 0 vpt2 V
  hpt2 neg 0 V Opaque stroke } def
/TriUW { stroke [] 0 setdash vpt 1.12 mul add M
  hpt neg vpt -1.62 mul V
  hpt 2 mul 0 V
  hpt neg vpt 1.62 mul V Opaque stroke } def
/TriDW { stroke [] 0 setdash vpt 1.12 mul sub M
  hpt neg vpt 1.62 mul V
  hpt 2 mul 0 V
  hpt neg vpt -1.62 mul V Opaque stroke } def
/PentW { stroke [] 0 setdash gsave
  translate 0 hpt M 4 {72 rotate 0 hpt L} repeat
  Opaque stroke grestore } def
/CircW { stroke [] 0 setdash 
  hpt 0 360 arc Opaque stroke } def
/BoxFill { gsave Rec 1 setgray fill grestore } def
end
}}%
\begin{picture}(3600,2160)(0,0)%
{\GNUPLOTspecial{"
gnudict begin
gsave
0 0 translate
0.100 0.100 scale
0 setgray
newpath
1.000 UL
LTb
400 300 M
63 0 V
2987 0 R
-63 0 V
400 551 M
63 0 V
2987 0 R
-63 0 V
400 803 M
63 0 V
2987 0 R
-63 0 V
400 1054 M
63 0 V
2987 0 R
-63 0 V
400 1306 M
63 0 V
2987 0 R
-63 0 V
400 1557 M
63 0 V
2987 0 R
-63 0 V
400 1809 M
63 0 V
2987 0 R
-63 0 V
400 2060 M
63 0 V
2987 0 R
-63 0 V
400 300 M
0 63 V
0 1697 R
0 -63 V
987 300 M
0 63 V
0 1697 R
0 -63 V
1573 300 M
0 63 V
0 1697 R
0 -63 V
2160 300 M
0 63 V
0 1697 R
0 -63 V
2746 300 M
0 63 V
0 1697 R
0 -63 V
3333 300 M
0 63 V
0 1697 R
0 -63 V
1.000 UL
LTb
400 300 M
3050 0 V
0 1760 V
-3050 0 V
400 300 L
1.000 UL
LTb
693 1683 M
-58 -98 V
9 32 R
-9 -32 V
24 23 V
1.000 UL
LTb
693 1255 M
0 125 V
9 -37 R
-9 37 V
-10 -37 V
1.000 UL
LT0
471 2060 M
1 -5 V
7 -35 V
7 -33 V
6 -30 V
7 -28 V
7 -26 V
7 -25 V
6 -23 V
7 -22 V
7 -21 V
6 -20 V
7 -18 V
7 -18 V
7 -18 V
6 -16 V
7 -16 V
7 -15 V
7 -15 V
6 -14 V
7 -14 V
7 -13 V
6 -13 V
7 -12 V
7 -12 V
7 -12 V
6 -11 V
7 -11 V
7 -11 V
7 -10 V
6 -10 V
7 -10 V
7 -10 V
7 -10 V
6 -9 V
7 -9 V
7 -9 V
6 -8 V
7 -9 V
7 -8 V
7 -8 V
6 -8 V
7 -8 V
7 -7 V
7 -8 V
6 -7 V
7 -8 V
7 -7 V
7 -7 V
6 -6 V
7 -7 V
7 -7 V
6 -6 V
7 -7 V
7 -6 V
7 -6 V
6 -6 V
7 -6 V
7 -6 V
7 -6 V
6 -6 V
7 -6 V
7 -5 V
7 -6 V
6 -5 V
7 -6 V
7 -5 V
6 -5 V
7 -5 V
7 -5 V
7 -6 V
6 -5 V
7 -4 V
7 -5 V
7 -5 V
6 -5 V
7 -4 V
7 -5 V
6 -5 V
7 -4 V
7 -5 V
7 -4 V
6 -4 V
7 -5 V
7 -4 V
7 -4 V
6 -5 V
7 -4 V
7 -4 V
7 -4 V
6 -4 V
7 -4 V
7 -4 V
6 -4 V
7 -4 V
7 -3 V
7 -4 V
6 -4 V
7 -4 V
1.000 UL
LT0
400 1557 M
20 -12 V
19 -12 V
20 -12 V
20 -12 V
20 -12 V
19 -12 V
20 -12 V
20 -12 V
20 -12 V
19 -12 V
20 -12 V
20 -12 V
20 -11 V
19 -12 V
20 -12 V
20 -11 V
20 -12 V
19 -11 V
20 -12 V
20 -11 V
20 -12 V
19 -11 V
20 -11 V
20 -12 V
20 -11 V
19 -11 V
20 -11 V
20 -12 V
20 -11 V
19 -11 V
20 -11 V
20 -11 V
20 -11 V
19 -11 V
20 -11 V
20 -11 V
20 -11 V
19 -11 V
20 -10 V
20 -11 V
20 -11 V
19 -11 V
20 -10 V
20 -11 V
20 -11 V
19 -10 V
20 -11 V
20 -10 V
20 -11 V
19 -10 V
20 -11 V
20 -10 V
20 -10 V
19 -11 V
20 -10 V
20 -10 V
20 -11 V
19 -10 V
20 -10 V
20 -10 V
20 -10 V
19 -11 V
20 -10 V
20 -10 V
20 -10 V
19 -10 V
20 -10 V
20 -10 V
20 -10 V
19 -10 V
20 -9 V
20 -10 V
20 -10 V
19 -10 V
20 -10 V
20 -9 V
20 -10 V
19 -10 V
20 -9 V
20 -10 V
20 -10 V
19 -9 V
20 -10 V
20 -9 V
20 -10 V
19 -9 V
20 -10 V
20 -9 V
20 -10 V
19 -9 V
20 -9 V
20 -10 V
20 -9 V
19 -9 V
20 -10 V
20 -9 V
20 -9 V
19 -9 V
20 -9 V
1.000 UL
LT0
2355 514 M
10 -5 V
10 -4 V
10 -5 V
10 -4 V
10 -5 V
9 -4 V
10 -4 V
10 -4 V
10 -4 V
10 -4 V
10 -4 V
10 -4 V
9 -4 V
10 -4 V
10 -4 V
10 -4 V
10 -3 V
10 -4 V
10 -3 V
10 -4 V
9 -3 V
10 -4 V
10 -3 V
10 -3 V
10 -4 V
10 -3 V
10 -3 V
10 -3 V
9 -3 V
10 -3 V
10 -3 V
10 -3 V
10 -3 V
10 -3 V
10 -3 V
10 -3 V
9 -2 V
10 -3 V
10 -2 V
10 -3 V
10 -2 V
10 -3 V
10 -2 V
10 -3 V
9 -2 V
10 -2 V
10 -3 V
10 -2 V
10 -2 V
10 -2 V
10 -2 V
10 -2 V
9 -2 V
10 -2 V
10 -2 V
10 -2 V
10 -1 V
10 -2 V
10 -2 V
10 -1 V
9 -2 V
10 -2 V
10 -1 V
10 -2 V
10 -1 V
10 -1 V
10 -2 V
10 -1 V
9 -1 V
10 -2 V
10 -1 V
10 -1 V
10 -1 V
10 -1 V
10 -1 V
10 -1 V
9 -1 V
10 -1 V
10 -1 V
10 -1 V
10 0 V
10 -1 V
10 -1 V
10 0 V
9 -1 V
10 -1 V
10 0 V
10 -1 V
10 0 V
10 0 V
10 -1 V
10 0 V
9 0 V
10 0 V
10 -1 V
10 0 V
10 0 V
10 0 V
10 0 V
0.600 UP
1.000 UL
LT6
1125 1134 CircleF
1125 1134 CircleF
1125 1134 CircleF
1125 1134 CircleF
1125 1134 CircleF
1125 1134 CircleF
1125 1134 CircleF
1125 1134 CircleF
1125 1134 CircleF
1125 1134 CircleF
1125 1134 CircleF
1125 1134 CircleF
1125 1134 CircleF
1125 1134 CircleF
1125 1134 CircleF
1125 1134 CircleF
1125 1134 CircleF
1125 1134 CircleF
1125 1134 CircleF
1125 1134 CircleF
1125 1134 CircleF
1125 1134 CircleF
1125 1134 CircleF
1125 1134 CircleF
1125 1134 CircleF
1125 1134 CircleF
1125 1134 CircleF
1125 1134 CircleF
1125 1134 CircleF
1125 1134 CircleF
1125 1134 CircleF
1125 1134 CircleF
1125 1134 CircleF
1125 1134 CircleF
1125 1134 CircleF
1125 1134 CircleF
1125 1134 CircleF
1125 1134 CircleF
1125 1134 CircleF
1125 1134 CircleF
1125 1134 CircleF
1125 1134 CircleF
1125 1134 CircleF
1125 1134 CircleF
1125 1134 CircleF
1125 1134 CircleF
1125 1134 CircleF
1125 1134 CircleF
1125 1134 CircleF
1125 1134 CircleF
1125 1134 CircleF
1125 1134 CircleF
1125 1134 CircleF
1125 1134 CircleF
1125 1134 CircleF
1125 1134 CircleF
1125 1134 CircleF
1125 1134 CircleF
1125 1134 CircleF
1125 1134 CircleF
1125 1134 CircleF
1125 1134 CircleF
1125 1134 CircleF
1125 1134 CircleF
1125 1134 CircleF
1125 1134 CircleF
1125 1134 CircleF
1125 1134 CircleF
1125 1134 CircleF
1125 1134 CircleF
1125 1134 CircleF
1125 1134 CircleF
1125 1134 CircleF
1125 1134 CircleF
1125 1134 CircleF
1125 1134 CircleF
1125 1134 CircleF
1125 1134 CircleF
1125 1134 CircleF
1125 1134 CircleF
1125 1134 CircleF
1125 1134 CircleF
1125 1134 CircleF
1125 1134 CircleF
1125 1134 CircleF
1125 1134 CircleF
1125 1134 CircleF
1125 1134 CircleF
1125 1134 CircleF
1125 1134 CircleF
1125 1134 CircleF
1125 1134 CircleF
1125 1134 CircleF
1125 1134 CircleF
1125 1134 CircleF
1125 1134 CircleF
1125 1134 CircleF
1125 1134 CircleF
1125 1134 CircleF
1125 1134 CircleF
0.600 UP
1.000 UL
LT6
2355 514 CircleF
2355 514 CircleF
2355 514 CircleF
2355 514 CircleF
2355 514 CircleF
2355 514 CircleF
2355 514 CircleF
2355 514 CircleF
2355 514 CircleF
2355 514 CircleF
2355 514 CircleF
2355 514 CircleF
2355 514 CircleF
2355 514 CircleF
2355 514 CircleF
2355 514 CircleF
2355 514 CircleF
2355 514 CircleF
2355 514 CircleF
2355 514 CircleF
2355 514 CircleF
2355 514 CircleF
2355 514 CircleF
2355 514 CircleF
2355 514 CircleF
2355 514 CircleF
2355 514 CircleF
2355 514 CircleF
2355 514 CircleF
2355 514 CircleF
2355 514 CircleF
2355 514 CircleF
2355 514 CircleF
2355 514 CircleF
2355 514 CircleF
2355 514 CircleF
2355 514 CircleF
2355 514 CircleF
2355 514 CircleF
2355 514 CircleF
2355 514 CircleF
2355 514 CircleF
2355 514 CircleF
2355 514 CircleF
2355 514 CircleF
2355 514 CircleF
2355 514 CircleF
2355 514 CircleF
2355 514 CircleF
2355 514 CircleF
2355 514 CircleF
2355 514 CircleF
2355 514 CircleF
2355 514 CircleF
2355 514 CircleF
2355 514 CircleF
2355 514 CircleF
2355 514 CircleF
2355 514 CircleF
2355 514 CircleF
2355 514 CircleF
2355 514 CircleF
2355 514 CircleF
2355 514 CircleF
2355 514 CircleF
2355 514 CircleF
2355 514 CircleF
2355 514 CircleF
2355 514 CircleF
2355 514 CircleF
2355 514 CircleF
2355 514 CircleF
2355 514 CircleF
2355 514 CircleF
2355 514 CircleF
2355 514 CircleF
2355 514 CircleF
2355 514 CircleF
2355 514 CircleF
2355 514 CircleF
2355 514 CircleF
2355 514 CircleF
2355 514 CircleF
2355 514 CircleF
2355 514 CircleF
2355 514 CircleF
2355 514 CircleF
2355 514 CircleF
2355 514 CircleF
2355 514 CircleF
2355 514 CircleF
2355 514 CircleF
2355 514 CircleF
2355 514 CircleF
2355 514 CircleF
2355 514 CircleF
2355 514 CircleF
2355 514 CircleF
2355 514 CircleF
2355 514 CircleF
0.600 UP
1.000 UL
LT6
3333 300 CircleF
3333 300 CircleF
3333 300 CircleF
3333 300 CircleF
3333 300 CircleF
3333 300 CircleF
3333 300 CircleF
3333 300 CircleF
3333 300 CircleF
3333 300 CircleF
3333 300 CircleF
3333 300 CircleF
3333 300 CircleF
3333 300 CircleF
3333 300 CircleF
3333 300 CircleF
3333 300 CircleF
3333 300 CircleF
3333 300 CircleF
3333 300 CircleF
3333 300 CircleF
3333 300 CircleF
3333 300 CircleF
3333 300 CircleF
3333 300 CircleF
3333 300 CircleF
3333 300 CircleF
3333 300 CircleF
3333 300 CircleF
3333 300 CircleF
3333 300 CircleF
3333 300 CircleF
3333 300 CircleF
3333 300 CircleF
3333 300 CircleF
3333 300 CircleF
3333 300 CircleF
3333 300 CircleF
3333 300 CircleF
3333 300 CircleF
3333 300 CircleF
3333 300 CircleF
3333 300 CircleF
3333 300 CircleF
3333 300 CircleF
3333 300 CircleF
3333 300 CircleF
3333 300 CircleF
3333 300 CircleF
3333 300 CircleF
3333 300 CircleF
3333 300 CircleF
3333 300 CircleF
3333 300 CircleF
3333 300 CircleF
3333 300 CircleF
3333 300 CircleF
3333 300 CircleF
3333 300 CircleF
3333 300 CircleF
3333 300 CircleF
3333 300 CircleF
3333 300 CircleF
3333 300 CircleF
3333 300 CircleF
3333 300 CircleF
3333 300 CircleF
3333 300 CircleF
3333 300 CircleF
3333 300 CircleF
3333 300 CircleF
3333 300 CircleF
3333 300 CircleF
3333 300 CircleF
3333 300 CircleF
3333 300 CircleF
3333 300 CircleF
3333 300 CircleF
3333 300 CircleF
3333 300 CircleF
3333 300 CircleF
3333 300 CircleF
3333 300 CircleF
3333 300 CircleF
3333 300 CircleF
3333 300 CircleF
3333 300 CircleF
3333 300 CircleF
3333 300 CircleF
3333 300 CircleF
3333 300 CircleF
3333 300 CircleF
3333 300 CircleF
3333 300 CircleF
3333 300 CircleF
3333 300 CircleF
3333 300 CircleF
3333 300 CircleF
3333 300 CircleF
3333 300 CircleF
1.000 UL
LTb
1768 1121 M
63 0 V
1367 0 R
-63 0 V
1768 1240 M
63 0 V
1367 0 R
-63 0 V
1768 1358 M
63 0 V
1367 0 R
-63 0 V
1768 1477 M
63 0 V
1367 0 R
-63 0 V
1768 1596 M
63 0 V
1367 0 R
-63 0 V
1768 1715 M
63 0 V
1367 0 R
-63 0 V
1768 1833 M
63 0 V
1367 0 R
-63 0 V
1768 1952 M
63 0 V
1367 0 R
-63 0 V
1768 1121 M
0 63 V
0 768 R
0 -63 V
231 -768 R
0 63 V
0 768 R
0 -63 V
230 -768 R
0 63 V
0 768 R
0 -63 V
231 -768 R
0 63 V
0 768 R
0 -63 V
231 -768 R
0 63 V
0 768 R
0 -63 V
230 -768 R
0 63 V
0 768 R
0 -63 V
231 -768 R
0 63 V
0 768 R
0 -63 V
1.000 UL
LTb
1768 1121 M
1430 0 V
0 831 V
-1430 0 V
0 -831 V
1.000 UL
LTb
1883 1833 M
-38 -95 V
3 30 R
-3 -30 V
18 24 V
1.000 UL
LTb
1883 1572 M
0 101 V
7 -30 R
-7 30 V
-8 -30 V
1.000 UL
LT0
2084 1569 M
-7 4 V
-6 3 V
-7 4 V
-6 3 V
-6 3 V
-7 4 V
-6 3 V
-6 4 V
-6 3 V
-6 3 V
-5 4 V
-6 3 V
-6 4 V
-5 3 V
-6 3 V
-5 4 V
-6 3 V
-5 4 V
-5 3 V
-5 3 V
-6 4 V
-5 3 V
-5 4 V
-4 3 V
-5 3 V
-5 4 V
-5 3 V
-4 3 V
-5 4 V
-4 3 V
-5 4 V
-4 3 V
-4 3 V
-4 4 V
-5 3 V
-4 4 V
-4 3 V
-4 3 V
-3 4 V
-4 3 V
-4 4 V
-4 3 V
-3 3 V
-4 4 V
-3 3 V
-4 4 V
-3 3 V
-3 3 V
-3 4 V
-4 3 V
-3 4 V
-3 3 V
-3 3 V
-2 4 V
-3 3 V
-3 4 V
-3 3 V
-2 3 V
-3 4 V
-2 3 V
-3 4 V
-2 3 V
-2 3 V
-3 4 V
-2 3 V
-2 4 V
-2 3 V
-2 3 V
-2 4 V
-2 3 V
-1 4 V
-2 3 V
-2 3 V
-1 4 V
-2 3 V
-1 4 V
-2 3 V
-1 3 V
-2 4 V
-1 3 V
-1 4 V
-1 3 V
-1 3 V
-1 4 V
-1 3 V
-1 4 V
-1 3 V
0 3 V
-1 4 V
0 3 V
-1 3 V
0 4 V
-1 3 V
0 4 V
-1 3 V
0 3 V
0 4 V
0 3 V
0 4 V
1.000 UL
LT0
2756 1223 M
-22 12 V
-22 11 V
-22 11 V
-22 11 V
-21 11 V
-20 11 V
-21 10 V
-20 11 V
-20 10 V
-19 10 V
-19 10 V
-19 9 V
-19 10 V
-18 9 V
-18 10 V
-17 9 V
-17 9 V
-18 8 V
-16 9 V
-17 8 V
-16 9 V
-16 8 V
-16 8 V
-15 8 V
-15 8 V
-15 7 V
-15 8 V
-14 7 V
-14 8 V
-14 7 V
-14 7 V
-13 7 V
-13 6 V
-13 7 V
-13 7 V
-12 6 V
-12 6 V
-12 6 V
-12 6 V
-12 6 V
-11 6 V
-11 6 V
-11 6 V
-11 5 V
-10 5 V
-10 6 V
-10 5 V
-10 5 V
-10 5 V
-9 5 V
-9 4 V
-9 5 V
-9 4 V
-8 5 V
-9 4 V
-8 4 V
-8 4 V
-8 4 V
-7 4 V
-8 4 V
-7 4 V
-7 3 V
-6 4 V
-7 3 V
-6 4 V
-7 3 V
-6 3 V
-5 3 V
-6 3 V
-5 2 V
-6 3 V
-5 3 V
-5 2 V
-4 3 V
-5 2 V
-4 2 V
-4 2 V
-4 2 V
-4 2 V
-3 2 V
-4 2 V
-3 1 V
-3 2 V
-3 1 V
-2 2 V
-3 1 V
-2 1 V
-2 1 V
-2 1 V
-2 1 V
-1 1 V
-2 1 V
-1 0 V
-1 1 V
-1 0 V
-1 1 V
1.000 UL
LT0
3152 1121 M
-5 0 V
-4 0 V
-5 0 V
-4 0 V
-5 0 V
-4 0 V
-5 1 V
-4 0 V
-5 0 V
-4 0 V
-5 1 V
-4 0 V
-5 0 V
-4 1 V
-4 0 V
-5 0 V
-4 1 V
-5 0 V
-4 1 V
-4 0 V
-5 1 V
-4 0 V
-4 1 V
-4 0 V
-5 1 V
-4 1 V
-4 0 V
-4 1 V
-5 1 V
-4 0 V
-4 1 V
-4 1 V
-4 1 V
-5 1 V
-4 0 V
-4 1 V
-4 1 V
-4 1 V
-4 1 V
-4 1 V
-4 1 V
-4 1 V
-4 1 V
-4 1 V
-4 1 V
-5 1 V
-4 1 V
-4 1 V
-3 1 V
-4 1 V
-4 1 V
-4 1 V
-4 1 V
-4 1 V
-4 2 V
-4 1 V
-4 1 V
-4 1 V
-4 2 V
-4 1 V
-3 1 V
-4 1 V
-4 2 V
-4 1 V
-4 1 V
-4 2 V
-3 1 V
-4 1 V
-4 2 V
-4 1 V
-3 2 V
-4 1 V
-4 2 V
-4 1 V
-3 2 V
-4 1 V
-4 2 V
-3 1 V
-4 2 V
-4 1 V
-3 2 V
-4 1 V
-4 2 V
-3 2 V
-4 1 V
-4 2 V
-3 1 V
-4 2 V
-3 2 V
-4 1 V
-3 2 V
-4 2 V
-3 2 V
-4 1 V
-4 2 V
-3 2 V
-4 2 V
-3 1 V
-4 2 V
0.600 UP
1.000 UL
LT6
2084 1569 CircleF
2084 1569 CircleF
2084 1569 CircleF
2084 1569 CircleF
2084 1569 CircleF
2084 1569 CircleF
2084 1569 CircleF
2084 1569 CircleF
2084 1569 CircleF
2084 1569 CircleF
2084 1569 CircleF
2084 1569 CircleF
2084 1569 CircleF
2084 1569 CircleF
2084 1569 CircleF
2084 1569 CircleF
2084 1569 CircleF
2084 1569 CircleF
2084 1569 CircleF
2084 1569 CircleF
2084 1569 CircleF
2084 1569 CircleF
2084 1569 CircleF
2084 1569 CircleF
2084 1569 CircleF
2084 1569 CircleF
2084 1569 CircleF
2084 1569 CircleF
2084 1569 CircleF
2084 1569 CircleF
2084 1569 CircleF
2084 1569 CircleF
2084 1569 CircleF
2084 1569 CircleF
2084 1569 CircleF
2084 1569 CircleF
2084 1569 CircleF
2084 1569 CircleF
2084 1569 CircleF
2084 1569 CircleF
2084 1569 CircleF
2084 1569 CircleF
2084 1569 CircleF
2084 1569 CircleF
2084 1569 CircleF
2084 1569 CircleF
2084 1569 CircleF
2084 1569 CircleF
2084 1569 CircleF
2084 1569 CircleF
2084 1569 CircleF
2084 1569 CircleF
2084 1569 CircleF
2084 1569 CircleF
2084 1569 CircleF
2084 1569 CircleF
2084 1569 CircleF
2084 1569 CircleF
2084 1569 CircleF
2084 1569 CircleF
2084 1569 CircleF
2084 1569 CircleF
2084 1569 CircleF
2084 1569 CircleF
2084 1569 CircleF
2084 1569 CircleF
2084 1569 CircleF
2084 1569 CircleF
2084 1569 CircleF
2084 1569 CircleF
2084 1569 CircleF
2084 1569 CircleF
2084 1569 CircleF
2084 1569 CircleF
2084 1569 CircleF
2084 1569 CircleF
2084 1569 CircleF
2084 1569 CircleF
2084 1569 CircleF
2084 1569 CircleF
2084 1569 CircleF
2084 1569 CircleF
2084 1569 CircleF
2084 1569 CircleF
2084 1569 CircleF
2084 1569 CircleF
2084 1569 CircleF
2084 1569 CircleF
2084 1569 CircleF
2084 1569 CircleF
2084 1569 CircleF
2084 1569 CircleF
2084 1569 CircleF
2084 1569 CircleF
2084 1569 CircleF
2084 1569 CircleF
2084 1569 CircleF
2084 1569 CircleF
2084 1569 CircleF
2084 1569 CircleF
0.600 UP
1.000 UL
LT6
2756 1223 CircleF
2756 1223 CircleF
2756 1223 CircleF
2756 1223 CircleF
2756 1223 CircleF
2756 1223 CircleF
2756 1223 CircleF
2756 1223 CircleF
2756 1223 CircleF
2756 1223 CircleF
2756 1223 CircleF
2756 1223 CircleF
2756 1223 CircleF
2756 1223 CircleF
2756 1223 CircleF
2756 1223 CircleF
2756 1223 CircleF
2756 1223 CircleF
2756 1223 CircleF
2756 1223 CircleF
2756 1223 CircleF
2756 1223 CircleF
2756 1223 CircleF
2756 1223 CircleF
2756 1223 CircleF
2756 1223 CircleF
2756 1223 CircleF
2756 1223 CircleF
2756 1223 CircleF
2756 1223 CircleF
2756 1223 CircleF
2756 1223 CircleF
2756 1223 CircleF
2756 1223 CircleF
2756 1223 CircleF
2756 1223 CircleF
2756 1223 CircleF
2756 1223 CircleF
2756 1223 CircleF
2756 1223 CircleF
2756 1223 CircleF
2756 1223 CircleF
2756 1223 CircleF
2756 1223 CircleF
2756 1223 CircleF
2756 1223 CircleF
2756 1223 CircleF
2756 1223 CircleF
2756 1223 CircleF
2756 1223 CircleF
2756 1223 CircleF
2756 1223 CircleF
2756 1223 CircleF
2756 1223 CircleF
2756 1223 CircleF
2756 1223 CircleF
2756 1223 CircleF
2756 1223 CircleF
2756 1223 CircleF
2756 1223 CircleF
2756 1223 CircleF
2756 1223 CircleF
2756 1223 CircleF
2756 1223 CircleF
2756 1223 CircleF
2756 1223 CircleF
2756 1223 CircleF
2756 1223 CircleF
2756 1223 CircleF
2756 1223 CircleF
2756 1223 CircleF
2756 1223 CircleF
2756 1223 CircleF
2756 1223 CircleF
2756 1223 CircleF
2756 1223 CircleF
2756 1223 CircleF
2756 1223 CircleF
2756 1223 CircleF
2756 1223 CircleF
2756 1223 CircleF
2756 1223 CircleF
2756 1223 CircleF
2756 1223 CircleF
2756 1223 CircleF
2756 1223 CircleF
2756 1223 CircleF
2756 1223 CircleF
2756 1223 CircleF
2756 1223 CircleF
2756 1223 CircleF
2756 1223 CircleF
2756 1223 CircleF
2756 1223 CircleF
2756 1223 CircleF
2756 1223 CircleF
2756 1223 CircleF
2756 1223 CircleF
2756 1223 CircleF
2756 1223 CircleF
0.600 UP
1.000 UL
LT6
3152 1121 CircleF
stroke
grestore
end
showpage
}}%
\put(1837,1536){\makebox(0,0)[l]{$E_{\rm av}$}}%
\put(1906,1833){\makebox(0,0)[l]{$E_{\rm typ}$}}%
\put(2483,871){\makebox(0,0){$R$}}%
\put(1468,1536){%
\special{ps: gsave currentpoint currentpoint translate
270 rotate neg exch neg exch translate}%
\makebox(0,0)[b]{\shortstack{$E$}}%
\special{ps: currentpoint grestore moveto}%
}%
\put(3152,1021){\makebox(0,0){0.6}}%
\put(2921,1021){\makebox(0,0){0.5}}%
\put(2691,1021){\makebox(0,0){0.4}}%
\put(2460,1021){\makebox(0,0){0.3}}%
\put(2229,1021){\makebox(0,0){0.2}}%
\put(1999,1021){\makebox(0,0){0.1}}%
\put(1768,1021){\makebox(0,0){0}}%
\put(1718,1952){\makebox(0,0)[r]{0.7}}%
\put(1718,1833){\makebox(0,0)[r]{0.6}}%
\put(1718,1715){\makebox(0,0)[r]{0.5}}%
\put(1718,1596){\makebox(0,0)[r]{0.4}}%
\put(1718,1477){\makebox(0,0)[r]{0.3}}%
\put(1718,1358){\makebox(0,0)[r]{0.2}}%
\put(1718,1240){\makebox(0,0)[r]{0.1}}%
\put(1718,1121){\makebox(0,0)[r]{0}}%
\put(3274,426){\makebox(0,0)[l]{$p_c$}}%
\put(2394,627){\makebox(0,0)[l]{$p_e$}}%
\put(1045,1004){\makebox(0,0)[l]{$p_y$}}%
\put(635,1105){\makebox(0,0)[l]{$E_{\rm av}$}}%
\put(752,1683){\makebox(0,0)[l]{$E_{\rm typ}$}}%
\put(1925,50){\makebox(0,0){$p$}}%
\put(100,1180){%
\special{ps: gsave currentpoint currentpoint translate
270 rotate neg exch neg exch translate}%
\makebox(0,0)[b]{\shortstack{$E$}}%
\special{ps: currentpoint grestore moveto}%
}%
\put(3333,200){\makebox(0,0){0.5}}%
\put(2746,200){\makebox(0,0){0.4}}%
\put(2160,200){\makebox(0,0){0.3}}%
\put(1573,200){\makebox(0,0){0.2}}%
\put(987,200){\makebox(0,0){0.1}}%
\put(400,200){\makebox(0,0){0}}%
\put(350,2060){\makebox(0,0)[r]{0.7}}%
\put(350,1809){\makebox(0,0)[r]{0.6}}%
\put(350,1557){\makebox(0,0)[r]{0.5}}%
\put(350,1306){\makebox(0,0)[r]{0.4}}%
\put(350,1054){\makebox(0,0)[r]{0.3}}%
\put(350,803){\makebox(0,0)[r]{0.2}}%
\put(350,551){\makebox(0,0)[r]{0.1}}%
\put(350,300){\makebox(0,0)[r]{0}}%
\end{picture}%
\endgroup
 